\documentclass[a4paper,11pt]{article}
\pdfoutput=1 

\usepackage{jcappub,aas_macros}
\usepackage[T1]{fontenc} 

\makeatletter
\gdef\@fpheader{Prepared for submission to JCAP\hfill
  \normalfont\sffamily MIT-CTP/6110}
\makeatother

\newcommand{\M}[0]{\mathrm{Mpc}/h}
\newcommand{\iM}[0]{h/\mathrm{Mpc}}

\newcommand{\qv}{\mathbf{q}}

\newcommand{\be}{\begin{equation}}
\newcommand{\ee}{\end{equation}}
\newcommand{\beqa}{\begin{eqnarray}}
\newcommand{\eeqa}{\end{eqnarray}}

\newcommand{\bsm}{\begin{smallmatrix}}
\newcommand{\esm}{\end{smallmatrix}}

\newcommand{\HH}{{\cal H}}

\newcommand\x{{\bf x}}
\renewcommand\k{{\bf k}}
\renewcommand\v{{\bf v}}

\newcommand\q{{\bf q}}

\newcommand{\bseq}{\begin{subequations}}
\newcommand{\eseq}{\end{subequations}}

\renewcommand{\ln}{\mathop{\rm ln}\nolimits}

\renewcommand{\H}{\mathcal H}

\renewcommand{\k}{{\bf k}}

\newcommand{\z}{{\bf z}}

\usepackage{stackengine}
\usepackage{float}
\usepackage{amsmath,amssymb,bm}
\usepackage{tikz}
\usetikzlibrary{arrows.meta}
\usepackage{cancel}
\usepackage{xcolor}

\usepackage{tikz}
\usepackage{cancel}
\usetikzlibrary{arrows.meta}

\tikzset{
  leg/.style={-Latex, line width=0.8pt},
  outleg/.style={Latex-, line width=0.9pt},
  vnode/.style={circle, draw, line width=0.8pt, inner sep=2pt, fill=white},
  lab/.style={font=\small},
}

\title{Toward Precision Kinetic Sunyaev-Zel'dovich Cosmology---1. The 
Matter -
Momentum 
Bispectrum at One-Loop Order 
}
\author[a,b,1]{Ryan Raikman,\note{Corresponding author.}}
\author[a,c,d,e]{James M. Sullivan }
\author[a,c,d]{Mikhail M. Ivanov}
\affiliation[a]{Department of Physics, Massachusetts Institute of Technology,
77 Massachusetts Avenue Cambridge, MA 02139, USA}
\affiliation[b]{MIT Kavli Institute for Astrophysics and Space Research,
Massachusetts Institute of Technology, 77 Massachusetts Avenue Cambridge, MA 02139, USA}
\affiliation[c]{Center for Theoretical Physics – a Leinweber Institute,
Massachusetts Institute of Technology, Cambridge, MA 02139, USA}
\affiliation[d]{The NSF AI Institute for Artificial Intelligence and Fundamental Interactions, Cambridge, MA 02139, USA}
\affiliation[e]{Brinson Prize Fellow}

\emailAdd{rraikman@mit.edu}
\emailAdd{jms3@mit.edu}
\emailAdd{ivanov99@mit.edu}

\abstract{
The kinetic Sunyaev-Zeldovich (kSZ) effect is a sensitive probe of the free-electron momentum field, tapping into spatial correlations relevant for both large-scale cosmology and smaller-scale astrophysics. 
Precision kSZ measurements will provide a robust channel for constraining inflationary physics and equivalence principle violation, while simultaneously characterizing the small-scale gas distribution and associated baryonic feedback.
However, robust kSZ measurement for cosmological applications is challenging, as the primary source of large-scale information is convolved with unknown small-scale dynamics in a degenerate combination. 
In this paper we address this challenge
using the Effective Field Theory 
(EFT) framework that provides a principled
way of treating the effects of small-scale
dynamics on large scale-structure. 
Extracting the kSZ signal 
can be reformulated as
a measurement 
of the 
one-loop
galaxy-galaxy-electron momentum bispectrum. 
We 
compute here a 
pure dark matter 
version of this bispectrum
in EFT 
and compare it to simulations. 
We derive the EFT counterterms
that capture 
small-scale backreaction
and find that two of them 
acquire a transverse direction.
One transverse counterterm gets generated 
entirely by the vorticity
of the velocity field.
Our computation 
of the bispectrum
dipole agrees with simulation data to
$5\%$ 
up to $k_{\rm max}=0.23~\iM$, which is significantly larger than the reach of tree level theory, which breaks down beyond $k_{\rm max}=0.07~\iM$ (at redshift $z=0.5$).
For the bispectrum computed with the projected momentum field, relevant to measuring the kSZ field, we find a similar $k_{\rm max}$, and make the first detection of the transverse counterterms 
from simulations. Our results
suggest that 
one-loop bispectrum computations can play
an important role in 
advancing 
kSZ cosmology.
}

\begin{document}
\maketitle

\section{Introduction \label{sec:intro}}

From reionization onward, free electrons suffuse the Universe, tracing the large-scale structure (LSS) within it.
Backlit by the Cosmic Microwave Background (CMB), these electrons impart their momentum to microwave photons via scattering, forming an important secondary temperature anisotropy signal.
This signal, the kinetic Sunyaev Zel'dovich effect (kSZ), must be controlled for to precisely characterize those anisotropies sourced by the primary CMB, but it also forms a sensitive probe of LSS in its own right.

This kSZ effect was predicted in the seminal work of Ref.~\cite{SZ1972,SZ1972AAP}, and first measured much later by Ref.~\cite{Hand2012:ksz} in cross correlation with galaxy survey data.
Subsequent work has improved the signal-to-noise of the detection \cite{Schaan16:kSZ_act_vel_recon_BOSS,Soergel16:ksz_spt,DeBernardis17:BOSS_ACT_ksz,Sugiyama2018:ksz_xcorr,Chaves-Montero21:ksz_ang,Schaan21:ksz_boss,Calafut21:act_sdss_ksz,McCarthy25:kSZ_ACT_DESILRG,Li26:act_cmass_joint,Qu2026:ksz_xcorr_desidr2_actdr6,Hadzhiyska26:ksz_desidr_actdr6}, and, today, kSZ cross-correlation measurements with galaxy positions can be reliably used to extract constraints on cosmological processes including local primordial non-Gaussianity \cite{Krywonos24:unwise_planck_fnl,Hotinli25:fnl_act_desils,Lague25:ksz_fnl,McCarthy25:kSZ_cont_fnl_act_desilrg,Chaussidon26:ksz_desidr2_actdr6} and baryonic feedback \cite{Hadzhiyska25:feedback_act_desi_ksz,RiedGuachalla25:ksz_feedback_desi1act,Siegel26:ksz_joint_feedback,Siegel26:xray_ksz_lens_act,Hadzhiyska25:feedback_ksz_lensing}.
The improvement of these measurements with recent CMB data has been impressive, and as kSZ measurement precision improves, concomitantly, theoretical model precision must closely follow.

Modeling of the kSZ has evolved significantly since the original work of Sunyaev and Zel'dovich \cite{SZ1972}.
Following Ref.~\cite{Kaiser1984,OstrikerVishniac:1986}, Ref.~\cite{Vishniac:1987_kSZ} subsequently developed a perturbative calculation for the transverse electron momentum auto-power spectrum, begetting related perturbative calculations \cite[e.g.,][]{Hu1994:reion_ov,Hu_ksz:2000,MaFryksz:02,Castro2003:ov_bisp_trisp,Castro2004:err_ov_bisp_trisp,Munshi16:kszlate_power_qpar_qperp,2016:Alvarez_ksz,Park2016:connected_qperp_sim}, including those accounting for ``patchy'' kSZ from inhomogeneous reionization \cite[e.g.,][]{Jaffe1998:ksz,Haiman1999:reion_cmb,Gruzinov1998:reion,Efstathiou:reion,Dodelson_reion:95,Gnedin2001:reioncmb,McQuinn05:patchy_ksz,kglee09:patchy_via_ksz_subtraction,Mesinger12:reion_ksz,Munshi16:ksz_patchy_sep,Smith17:patchy_ksz,Farren22:ula_ksz_halo,Hotinli23:He_reion_patchy,Chen23:patchy_ksz_morph,MacCrann24:ksz_trisp_patchy,Kumar25:patchy_ksz,Kramer25:ksz_patchy_cross,Lopez26:patchy}.
More recent modeling developments include the introduction of various kSZ statistics (e.g. kSZ$^2$, stacking, projected-field estimators, velocity reconstruction), in particular in cross correlation with other surveys \cite{Dore:2004_kSZsq,Fosalba07:xvpec,Ho09:vel_recon_orig,Li14:ksz_vel_matched_filter,Ferraro2016:kSZsq_methods,Hill16:ksz_proj_fields,Alonso16:ksz_recon_forecast,Ma18:kszq_21cm,Deutsch18:recon_ksz_later,Zhu20:ksz_vel_recon_for_bao,Sato-Polito21:v_recon_lim,LaPlante22:kszsq_reion,Contreras23:ml_vel_recon,Bolliet2023:proj_field_ksz,Patki23:ksz_proj_model,Schutt24:patchy_screening_lss,Zhou25:kszsq_21cmsq,Harscouet2026:pseudocl_stacking,Patki26:gravity_ksz_cmbl,Patki25:bisp_ksz_proj,Adolff26:ksz_de,Kumar25:electron_trisp,shearkszb:2026,shearkszg:2026}.
These methods largely rely upon approximations to the kSZ signal from simulations, possibly with templates, or halo models \cite[e.g.,][]{Thomas89:gas_sim,Scaramella93:gas_sim, Persi95:gas_sim,daSilva00:gas_sim,Seljak01:kszsim,Springel2001ksz,Zhang04:hydro_ksz_sim,Zhang02hydro_ksz_sim,Iliev07:rt_ksz_patchy_sim,Trac2011:sz_templates_hydro,Battaglia10:ksz_sim,Shaw12:ksz_sim,Dolag16:magneticum,Sugiyama17:forecast_ksz_galaxy_xcorr,Madhavacheril19:ksz_frb,Stein20:websky,Osato23:bary_pasting,Bigwood26:feedback_sims_ksz,Lague26:baryon_flamingo_emulator_ksz,Rodriguez26:proj_ksz_xcorr_webskyvshalo}.
In particular, halo models have recently been used to provide a modeling framework for both the kSZ transverse momentum auto-power spectrum \cite{Wayland2026:stacking_halo,Giri_2022}
and the galaxy-galaxy-electron momentum bispectrum \cite{Smith26PhRvD:ksz_bisp}.

The focus of this paper is to instead work toward a robust description of the kSZ power spectrum and bispectrum in cross correlation with galaxies.
We achieve this by applying Effective Field Theory (EFT) \cite[e.g.,][]{McDonald:2009dh,Baumann:2010tm,Carrasco:2012cv,Desjacques:2016bnm,Ivanov22:eftlss_review} for LSS, heavily drawing on the work related to the EFT of velocities and redshift-space distortions \cite[e.g.,][]{Carrasco:2013mua,Mercolli_2014,Senatore:2014vja,Perko:2016puo,Vlah:2016bcl,Vlah:2018ygt,Desjacques:2018pfv,Ivanov:2019pdj,Chen:2020fxs,Chen:2020fxs,Chudaykin:2020hbf,Chen26:rsd_vpec}.
The anisotropies induced in the CMB temperature field can be written as 
\begin{align}\label{eq:ksz_integral_defn}
	\frac{\delta T_{\text{kSZ}}(\hat{\bm{n}})}{T_{\text{CMB}}}
	= -\sigma_T \, n_{e,0} \int_0^{z_{\text{re}}} \frac{dz}{H(z)} \;
	x_e(z)\, e^{-\tau(z)} (1+z)^2 \;
	\hat{\bm{n}} \cdot \bm{\pi}\big(\chi(z)\hat{\bm{n}};\, z\big),
\end{align}
where $\sigma_T$ is the Thomson cross-section, $n_{e, 0}$ is the present day electron density, $H(z)$ is the Hubble expansion factor, $x_e(z)$ is the electron ionization fraction, $\tau(z)$ is the optical depth, and $\bm{\pi}$ is the electron momentum field\footnote{Here we write the expression in an approximation most closely connected to the topic of this work, the EFT of the momentum field. 
However, both $x_e$ and $\tau$ should be promoted to spatially varying fields as well in full generality, and should be accompanied by their own perturbative expansions. 
Such a general expression also captures, e.g., the patchy kSZ effect from reionization. 
}. Modulo the factors which vary relatively slowly with redshift, the kSZ field can be simply written as the electron momentum field projected along the line of sight. Like Ref.~\cite{Smith26PhRvD:ksz_bisp}, we focus on the $\langle \delta_g \delta_g \pi_e^\parallel\rangle'$ 3-point function in Fourier space, shown to be equivalent to large-scale velocity reconstruction and several other popular kSZ statistics. While the kSZ field is sourced by the projected momentum field, in this paper we primarily consider the full 3-dimensional momentum field, enabling a significantly more stringent comparison of theoretical expressions against simulations. 

We develop an EFT model for the matter density-density-momentum bispectrum at one-loop order in real space, drawing on the structure developed for matter and galaxy correlations in real and redshift space \cite[e.g.,][]{Baldauf:2014qfa,Eggemeier:2018qae,Eggemeier:2021cam,Ivanov:2021kcd,Philcox:2021kcw,Philcox:2022frc,Ivanov:2023qzb,DAmico:2022ukl,bakx2025oneloopgalaxybispectrumconsistent}.
In addition, we carry out an 
efficient computation of our theory predictions using the FFTLog technique~\cite{Simonovi__2018,Philcox:2022frc}.
The restriction of our attention to dark matter statistics is in fact not much of a limitation, as the free electron density almost exactly traces that of the underlying dark matter \cite[e.g.,][]{Andrew2026:unbiased_e}, and, by the equivalence principle, the electron velocity must also follow that of the dark matter.
We test the model on the FLAMINGO simulations, a significant improvement over the tree-level model as well the model naively including nonlinearity by substituting $P_L \to P_{NL}$. Relevant to the EFT model are counterterm corrections, representing renormalization of diverging or identically cutoff-dependent quantities. We derive these corrections, finding unique transverse structure in the momentum field. In fitting these corrections via a comparison to simulation, we present the first detection of these transverse counterterms.

This paper is organized as follows. Following the introduction, we introduce the perturbation theory notation for the relevant fields --- $\delta_m$, the matter overdensity, and $\pi_i$, the $i$-th component of the (vector) momentum field ---and compute $P_{\delta\delta}(k)$, $P_{\delta \pi_z}(k)$, and $P_{\pi_z \pi_z}(k)$ power spectra within the EFT framework in Section~\ref{sec:pk}. In Section~\ref{sec:bispectrum}, we derive the 1-loop corrections to the $\langle \delta \delta \pi_z\rangle$ bispectrum. In Section~\ref{sec:fftlog}, we describe the FFTLog algorithm used to efficiently compute these 1-loop corrections and demonstrate its accuracy with a comparison to brute-force numerical integration. In Section~\ref{sec:uv}, we describe how counterterm corrections arise in the EFTofLSS framework to yield consistent divergence-free theoretical expressions. In Section~\ref{sec:binning}, we detail the computation of bin averaged bispectra multipoles from the FLAMINGO simulations, and how the theoretical expressions are analogously binned. In Section~\ref{sec:sims}, we present the comparison between the theoretical expressions against simulations at both the power spectrum and bispectrum levels. We finish with the Conclusion in Section~\ref{sec:conc}, and leave some calculations to the Appendix.

\section{One-Loop $\langle \delta \delta \rangle$, $\langle \delta \pi_z \rangle$, and $\langle \pi_z \pi_z \rangle$ Power Spectra \label{sec:pk}}
We begin by computing all auto- and cross- power spectra of these two fields. This serves as both a warm-up to the EFT formalism and a sanity check when compared against simulations. We begin with Standard Perturbation Theory (SPT), which computes a perturbative solution to the Euler-Vlasov system \cite{Bernardeau:2001qr,ivanov2022effectivefieldtheorylarge} (Equation \eqref{eq:exact}) under the assumption that dark matter is a perfect fluid (i.e. $\tau_{\text{vis}}=0$), allowing us to express corrections to any correlation function in terms of the linear matter power spectrum $P_{11}(k, z)$ and integrals over it. EFT corrects the errors 
produced by this perfect fluid 
description.
In particular, it renormalizes 
the errors 
in the SPT loop diagrams
with counterterms, which are cutoff-dependent Wilson coefficients that must be matched to the data. Importantly, we will see that the EFT corrections introduce new physical 
ingredients absent
in SPT --- a transverse 
velocity 
component 
at the one-loop bispectrum order.

\subsection{SPT Kernel Expressions}
We begin with the perturbative expansions for the matter overdensity field ($\delta_m$) and matter momentum field ($\pi^i$) in SPT. For the matter momentum field, we begin by transforming to Fourier space\footnote{We use the shorthand \be \int_\q \dots \equiv \int \frac{d^3\q}{(2\pi)^3}\ee}:
\begin{align}\label{eq:pi_x_fourier}
    \pi_i(\x) &= {v}_i(\x) (1+\delta_m(\x)) \notag\\
    \pi_i(\k) &= {v}_i(\k) + \int_{\q} {v}_i(\q) \delta_m(\k-\q) \notag\\
                &= \frac{i k_i}{k^2}f\HH\theta(\k)+ \int_{\q}\frac{iq_{ i}}{q^2}f\HH\theta(\q) \delta_m(\k-\q),
\end{align}
where we have introduced the (scaled) velocity divergence $\theta(\k)$, defined via 
\be 
\theta(\x) = -\partial^i v_i(\x)/(f\HH)\,,
\ee 
where $f$ is the logarithmic
growth factor and $\HH$ is the conformal Hubble parameter. 
As in the language of SPT, both $\delta(\k)$ and $\theta(\k)$ admit perturbative expansions in powers of the linear matter density field $\delta_1(\k)$ \cite{Bernardeau:2001qr,ivanov2022effectivefieldtheorylarge}. We write the overall fields as sums over these perturbative expansions, effectively defining the kernels $F_n$ and ${M}_n^i$:
\be 
\begin{split}
& \delta(\k)=\sum_{n=1}\int_{\q_1,...,\q_n}F_n(\q_1,...,\q_n)(2\pi)^3\delta_D^{(3)}
(\k-\q_{1...n})\prod^n_{i=1}\delta_{(1)}(\qv_i)\,,\\
&\pi_i(\k)=i(f\HH)\sum_{n=1}\int_{\q_1,...,\q_n}{M}^i_n(\q_1,...,\q_n)(2\pi)^3\delta_D^{(3)}
(\k-\q_{1...n})\prod^n_{i=1}\delta_{(1)}(\qv_i)\,,
\end{split}
\ee 
where $F_n$ is the usual density kernel \cite{ivanov2022effectivefieldtheorylarge}. From Eq. \eqref{eq:pi_x_fourier}, by using the perturbative expansion of $\delta(\k)$ and $\theta(\k)$ in terms of the kernels $F_n$ and $G_n$ respectively, we obtain the momentum field expansion kernels $M_n$:

\begin{align}\label{eq:M_n_sum}
     M^i_n(\bm{q_1}...\bm{q_n}) = if\HH\left[\frac{ {q}^i_{1...n}}{{q}_{1...n}^2} G_n(\bm{q}_1... \bm{q}_n) + \sum_{m=1}^{n-1} \frac{ {q}^i_{1...m}}{{q}_{1...m}^2} G_m(\bm{q}_1,... \bm{q}_m) F_{n-m}(\bm{q}_{m+1}...\bm{q}_n)\right],
\end{align}
where $\q_{1..m}\equiv \q_1+\q_2 +... \q_m$. The first term comes from the bare velocity field, and the second comes from the convolved velocity and density fields. The explicit forms for the unsymmetrized momentum field kernels $M_n$ are provided in Appendix \ref{app:M_n}. In writing the above perturbative expansions, we have used the Einstein-de Sitter approximation to factorize the time dependence from the perturbative kernels. Under this approximation, shown to induce error below the percent level \cite{Takahashi:2008yk,Fasiello:2016qpn,Fasiello:2022lff}, the $\delta^{(n)}$ has time dependence like $D^n(\tau)$, where $D(\tau)$ is the linear growth factor.

While $F_n$ and $G_n$ kernels are even under parity 
transformation $\q_i\to -\q_i$ for any $i$,
$M^i_n$ is odd w.r.t. an overall
inversion of its arguments, 
\be 
M_n^i(-\q_1,...,-\q_n)=-M^i_n(\q_1,...,\q_n)\,.
\ee 
In what follows we will also use the 
tilde momentum kernels without the 
factors of $if\HH$:
\be 
{M}^i_n \equiv if\HH \tilde{M}^i_n\,.
\ee 

\subsection{SPT Power Spectrum Expressions}
We proceed with the power spectra computations, beginning with the matter-matter power spectrum \cite{ivanov2022effectivefieldtheorylarge}. Here, our goal is to express the expectation of $\langle \delta(\k_1) \delta^*(\k_2)\rangle $ in terms of the linear matter power spectrum, defined via $\langle \delta_{(1)}(\k_1) \delta_{(1)}^*(\k_2)\rangle = (2 \pi)^3 \delta_D^{(3)}(\k_1+\k_2) P_{11}(k_1)$. We use the notation $\langle \delta_{(1)}(\k_1) \delta_{(1)}(\k_2)\rangle' = P_{11}(k_1)$, i.e. stripping the Dirac delta function. The density field perturbative correction at order $n$, $\delta_{(n)}$, contains $n$ powers of $\delta_{(1)}(\q)$. Since $\delta_{(1)}(\q)$ is a Gaussian random field, the expectation of odd powers of $\delta_{(1)}(\q)$, e.g. $\langle \delta_{(1)}(\k_1) \delta_{(1)}(\k_2) \delta_{(1)}(\k_3) \rangle $ is exactly zero, and the expectation of even powers can be decomposed into $P_{11}(k)$ and Dirac delta functions via a Wick expansion \cite{Bernardeau:2001qr, ivanov2022effectivefieldtheorylarge}. As such, we have the pure linear power spectrum at leading order, and at next to leading order (NLO), analogously referred to as 1-loop order due to the presence of the integral over $\q$, we have two contributions:
\begin{align}
    \langle \delta(\k_1) \delta(\k_2) \rangle' &= P_{11}(k) + 2\int_{\q} [F_2(\k_2-\q,\q)]^2 P_{11}(q)P_{11}(|\k_2-\q|) \notag\\
& \qquad +6 P_{11}(k)\int_\q F_3(\k_2,-\q,\q)P_{11}(q).
\end{align}
The term with $(F_2)^2$ is referred to as $P_{22}$, as it has both fields at second order. The term with $F_3$ is referred to as $P_{13}$, with one field at first order and the other at third order. We continue with the density-momentum power spectrum, following very similar structure but with the $M_n$ kernels from the momentum field:
\begin{align}
\langle 
&\pi_i(\k_1)\delta(\k_2) \rangle'=
if \HH \tilde{P}^i_{\delta \pi}(\k_1)\text{, where}\notag\\
 & \tilde{P}^i_{\delta \pi}(\k) = \frac{k_{i}}{k^2}P_{11}(k) + 2\int_{\q} F_2(\k-\q,\q)\tilde{M}^i_2(\k-\q,\q)P_{11}(q)P_{11}(|\k-\q|) \,,\notag\\ 
& \qquad +3 P_{11}(k)\left(\frac{k_{i}}{k^2}\int_\q F_3(\k,-\q,\q)P_{11}(q)+
F_1(\k_1)\int_\q \tilde{M}^i_3(-\k,-\q,\q)P_{11}(q)\right)\,.
\end{align}
 As we have two different fields, the $P_{13}$ term can involve either the density field or momentum field at third order. Finally, for the momentum-momentum power spectrum: 
\begin{align}\label{eq:Pi1l}
&\langle \pi_i(\bm{k}_1) \pi_j(\bm{k}_2) \rangle' = -(f\HH)^2 \tilde{P}^{ij}_{\pi\pi}(\k_1)\,\notag\\
  & \tilde{P}^{ij}_{\pi\pi}(\k) = \frac{k_{i} k_{ j}}{k^4}P_{11}(k) + 2\int_{\q} \tilde M^i_2(\k-\q,\q)
   \tilde M^j_2(\k-\q,\q)P_{11}(q)P_{11}(|\k-\q|) \notag\,,\\
& \qquad + 3P_{11}(k) \left[ \frac{k_{i}}{k^2}\int_\q \tilde{M}^j_3(\k,-\q,\q)P_{11}(q) + \frac{k_{j}}{k^2}\int_\q \tilde{M}^i_3(\k,-\q,\q)P_{11}(q)
\right]
\end{align}
 To compare this power spectrum against simulation, we compute it purely for the z-component of the momentum field, i.e. $\langle \pi_z(\k_1) \pi_z(\k_2) \rangle$. In computing theoretical predictions for the power spectra, we use $P_{11}$ computed by \texttt{camb} \cite{Lewis_2000} Einstein-Boltzmann solver for the D3A \cite{Schaye_2023, DES:2021wwk} cosmology used to run simulations, discussed further in Section~\ref{sec:sims}. To compute loop corrections, we use the FFTLog formalism \cite{Simonovi__2018, Hamilton:1999uv}, formally introduced in Section~\ref{sec:fftlog} for the more involved bispectrum loops. These loop corrections are computed with an IR-resummed power spectrum, described by \cite{ Blas:2016sfa, Ivanov:2018gjr} (see also~\cite{Senatore:2014via,Baldauf:2015xfa,Vlah:2015zda,Chen:2020fxs,Chen:2020zjt}). 

\section{The One-Loop $\langle \delta\,\delta\,\pi_z \rangle$ Bispectrum \label{sec:bispectrum}}
Having computed power spectra of the $\delta_m$ and $\pi_z$ fields in the above section to 1-loop order, we continue with the $\langle \delta\,\delta\,\pi_z \rangle$ bispectrum. We use the same perturbative expressions for our fields in terms of $F_n$ and $M^i_n$. To obtain a non-vanishing bispectrum prediction, we search for combinations of the three fields, each at different order in $\delta_{(1)}$, such that the overall power of $\delta_{(1)}$ is even. At the lowest order, or ``tree-level,'' we can have one field at second-order and the other at first, i.e. $\mathcal{O}(\delta_{(1)}^4) \sim \mathcal{O}(P_{11}^2)$. At NLO, or "1-loop", there are four ways to obtain $\mathcal{O}(\delta_{(1)}^6) \sim \mathcal{O}(P_{11}^3)$, denoted $B_{222}$, $B_{321}^I$, $B_{321}^{II}$, and $B_{411}$ \cite{Scoccimarro:1997st, Bernardeau:2001qr, Baldauf:2014qfa}. These graphs represent different topologies in computing the 1-loop correction, involving different orders of perturbation theory kernels and ways of connecting internal lines. The structure of these graphs is identical to other cases of the 1-loop bispectrum, for example $\langle \delta \delta \delta \rangle$, but the presence of a differentiated momentum field induces different permutations in each graph.

\subsection{Bispectrum Loop Corrections}

We begin by writing the SPT bispectrum as the sum of terms through the one-loop order:
\begin{align}
	\langle\delta(-\k_1) \delta(-\k_2) \pi_z(-\k_3) \rangle' 
    &= i f \HH \tilde{B}_{\delta \delta \pi_z}(\k_1, \k_2, \k_3)\notag \\&=i f \HH(B_{211}^z(\k_1, \k_2, \k_3) + B_{222}^z(\k_1, \k_2, \k_3) \notag\\ 
	&\quad + B_{321}^{I, z}(\k_1, \k_2, \k_3) + B_{321}^{II, z}(\k_1, \k_2, \k_3) + B_{411}^z(\k_1, \k_2, \k_3)) \,.
\end{align}
 The tree-level kSZ bispectrum was computed by \cite{Smith26PhRvD:ksz_bisp}.
In our expression we generalize it to an unprojected momentum field (i.e. not specialized to the 2-dimensional kSZ map) by including contributions with $k_{3,z} \neq 0$, such as $M_1(-\k_3)$. We also consistently extend this correlator beyond the tree order. 
For general momentum directions, we have:
\be 
\begin{split}
\langle \delta(-\k_1) \delta(-\k_2) \pi_i(-\k_3) \rangle &= i f \HH(B_{211}^i(\k_1, \k_2, \k_3) + B_{222}^i(\k_1, \k_2, \k_3) \notag\\ 
	&\quad + B_{321}^{I, i}(\k_1, \k_2, \k_3) + B_{321}^{II, i}(\k_1, \k_2, \k_3) + B_{411}^i(\k_1, \k_2, \k_3))\,.
    \end{split}
\ee 
 At the tree level, we have:
\begin{align}
	B^i_{211}(\k_1, \k_2, \k_3) &= 2 \tilde{M}^i_2(\k_1, \k_2) F_1(\k_1)F_1(\k_2)P_{11}(k_1)P_{11}(k_2) \notag \\
    &\quad + \left[2 F_2(\k_1, \k_3) F_1(\k_1) \tilde{M}^i_1(-\k_3) P_{11}(k_1)P_{11}(k_3) + \{ \k_1 \leftrightarrow \k_2\}\right].
\end{align}
We continue with the four 1-loop bispectrum corrections, representing all possible contractions of the three fields at order $\mathcal{O}(\delta^6)$. 
As mentioned earlier, unlike the $F_n$ kernels which are even under inversion of their arguments, the momentum field kernels $M_n$ are odd. For example, $F_2(-\bm{q_1}, -\bm{q_2})= F_2(\bm{q_1}, \bm{q_2})$, but $M^i_2(-\bm{q_1}, -\bm{q_2})= -M^i_2(\bm{q_1}, \bm{q_2})$. As a consequence, when writing the loop corrections, attention must be paid that the exact sign of arguments are consistent with the direction of momentum flow in the graphs. Explicitly, we have: 

\begin{align}\label{eq:bispec_loops}
	B_{222}^i &= 8\int_{\q}
   F_{2}(-\k_1-\q,\q)\,
   \tilde{M}_{2}^i(\k_1+\q,\k_2-\q)\,
   F_{2}(-\k_2+\q,-\q) P_{11}(q)\,
   P_{11}(|\k_1+\q|)\,
   P_{11}(|\k_2-\q|)\,\notag
   \\
	B_{321}^{I, i} &= 6 \, F_1(\k_1) P_{11}(k_1)
   \int_{\q} 
   \tilde{M}_3^i(\q, -\q+\k_2, \k_1)\,
   F_2(-\q, -\k_2+\q)\,
   P_{11}(q)\, P_{11}(|\k_2-\q|)+\{\k_1 \leftrightarrow \k_2\} \notag \\
   & \qquad + 6 \, F_1(\k_1) P_{11}(k_1)
   \int_{\q} 
   F_3(\q, -\q+\k_3, \k_1)\,
   \tilde{M}_2^i(-\q, -\k_3+\q) \,
   P_{11}(q)\, P_{11}(|\k_3-\q|) + \{\k_1 \leftrightarrow \k_2 \} \notag\\
   & \qquad + 6 \, \tilde{M}_1^i(-\k_3) P_{11}(k_3)
   \int_{\q} 
   F_3(\q, -\q+\k_1, \k_3)\,
   F_2(-\q, \q-\k_1) \,
   P_{11}(q)\, P_{11}(|\k_1-\q|) + \{\k_1 \leftrightarrow \k_2 \} \notag\\
	B_{321}^{II, i} &= 6 \, F_2(\k_3,\k_1)\,
    F_1(\k_1)\,
   P_{11}(k_3)\, P_{11}(k_1)
   \int_{\q}
   \tilde{M}_3^i(-\k_3,\q, -\q)\,
   P_{11}(q) + \{\k_1 \leftrightarrow \k_2\} \notag \\
   & \qquad +6 \, \tilde{M}_2^i(\k_1,\k_2)\,
    F_1(\k_1)\,
   P_{11}(k_1)\, P_{11}(k_2)
   \int_{\q}
   F_3(-\k_2,\q, -\q)\,
   P_{11}(q) + \{\k_1 \leftrightarrow \k_2\}\notag \\
   & \qquad +6 \, F_2(\k_1,\k_3)\,
    \tilde{M}_1^i(-\k_3)\,
   P_{11}(k_1)\, P_{11}(k_3)
   \int_{\q}
   F_3(-\k_1,\q, -\q)\,
   P_{11}(q) + \{\k_1 \leftrightarrow \k_2\}\notag\\
	B_{411}^i &= 12 \, F_1(\k_1)\, F_1(\k_2)\,
   P_{11}(k_1)\, P_{11}(k_2)
   \int_{\q}
\tilde{M}_4^i(\k_1,\k_2,\q, -\q)\,
   P_{11}(q)\notag\\
   &\qquad+ 12 \, F_1(\k_1)\, \tilde{M}_1^i(-\k_3)\,
   P_{11}(k_1)\, P_{11}(k_3)
   \int_{\q}
F_4(\k_1,\k_3,\q, -\q)\,
   P_{11}(q)+ \{\k_1 \leftrightarrow \k_2\}.
\end{align}
To compute these 1-loop corrections, one can proceed directly with numerical integration, being careful about the dependence of results on the limits of integration chosen. However, this is prohibitively slow when evaluating the number of triangle configurations needed for a comprehensive comparison to simulation.

\section{Efficient Evaluation with FFTLog \label{sec:fftlog}}

In this section, we provide a brief overview of the FFTLog formalism used to efficiently compute 1-loop integrals in Eqs. \eqref{eq:bispec_loops}. For a more detailed description, we refer the reader to \cite{Simonovi__2018,Chudaykin:2020aoj,Philcox:2022frc}. Especially for the bispectrum, where loop corrections must be evaluated over many triangle configurations $k_1, k_2, k_3$, brute-force numerical integration is prohibitively expensive and slow. FFTLog provides a way to compute these loop corrections by decomposing the integrand into polynomials, which can then be evaluated as a sum over hypergeometric functions. Within the integrand we have both perturbation theory kernels, which we can straightforwardly re-express in terms of polynomials, and the linear theory power spectra $P_{11}(q)$, which in principle are arbitrary functions. The backbone of FFTLog is to express $P_{11}(q)$ as a finite sum over polynomials with complex exponents, enabling efficient computation of the loop corrections.

\subsection{FFTLog Algorithm}
To detail the FFTLog algorithm, we use the $B_{222}$ loop as an example, as the formalism straightforwardly generalizes to the other loop corrections in Eqs. \eqref{eq:bispec_loops}. Our presentation is very similar to \cite{Simonovi__2018} with the inclusion of the line of sight dependent kernels. These have been thoroughly developed for the case of galaxy redshift-space distortions \cite{Philcox:2022frc, DAmico:2022ukl, bakx2025oneloopgalaxybispectrumconsistent, Ivanov:2021kcd}, and we present a brief overview. To summarize, we are interested in evaluating an integral like:
\begin{align}\label{eq:b222_fftlog_ex}
    B_{222}^z(\k_1, \k_2, \k_3) &= 8\int_{\q}
   F_{2}(-\k_1-\q,\q)\,
   \tilde{M}^z_{2}(\k_1+\q,\k_2-\q)\,
   F_{2}(-\k_2+\q,-\q)\notag\\ & \qquad \qquad \times P_{11}(q)\,
   P_{11}(|\k_1+\q|)\,
   P_{11}(|\k_2-\q|),
\end{align}
with the integral taken over all $\q$ for fixed values of $\k_1, \k_2, \k_3$. Our goal is to express each component of the integrand as a sum over polynomials. Once this is accomplished, we can use the integral relation \cite{Simonovi__2018}:
\begin{align}\label{eq:fftlog_J_defn}
    \int_\q \frac{1}{q^{2\nu_1} |\k_1-\q|^{2\nu_2} |\k_2+\q|^{2\mu_3}}\equiv k_1^{3-2\nu_1-2\nu_2-2\nu_3} J\left(\nu_1, \nu_2, \nu_3; x, y\right),
\end{align}
 where we have defined $x\equiv k_3^2/k_1^2, y\equiv k_2^2/k_1^2$, and $J$ can be expressed in terms of hypergeometric functions, see Equation 3.32 in \cite{Simonovi__2018}. Starting with the linear matter power spectrum $P_{11}(q)$, the standard FFTLog decomposition is given by:
\begin{align}\label{eq:fftlog_p11_decomp}
    P_{11}(q) = \sum_{m=1}^{N_m} c_m q^{\nu + i \eta_m},
\end{align}
where $\nu$ is the (real) FFTLog "bias", and the decomposition is for $N_m$ complex frequencies $\eta_m$ with coefficients $c_m$. For logarithmically spaced frequencies $\eta_m$, these coefficients are computed to best match the input linear power spectrum. The FFTLog "bias" $\nu$ allows the computation to avoid certain divergences (either UV or IR), as non-convergent integrals are zeroed in the FFTLog formalism. In all of our computations we use $\nu=-1.6$, which retains the UV parts of the integrals, 
and sets to zero the leading IR contributors (sometimes called IR enhancements or IR divergences~\cite{Blas:2015qsi}), which cancel 
in our bispectrum anyway. 
Note that this is consistent with the prescription used for the one-loop FFTLog computations of Lyman-$\alpha$
forest power spectrum~\cite{Ivanov:2023yla}.

Next, we must treat the perturbation theory kernels $F_n, M_n$. The matter kernel $F_2(-\k_2+\q, -\q)$ is a polynomial in possible dot products between the two vectors, therefore can be expressed in terms of integer powers of $k_2, q$, and $|\k_2-\q|$, given explicitly in Eq 2.4 by \cite{Simonovi__2018}. If only the matter kernels $F_n$ were present, we could decompose Eq. \eqref{eq:b222_fftlog_ex} into a sum over polynomials, schematically as:
\begin{align}\label{eq:fftlog_b222_explicit}
    &B_{222}(k_1, k_2, k_3) = \sum_{m_1, m_2, m_3}M_{222}(\nu_1, \nu_2, \nu_3; x, y)c_{m_1}c_{m_2}c_{m_3} k_1^{3-2\nu_1-2 \nu_2-2 \nu_3}, \text{ where}\notag\\
    &M_{222}(\nu_1, \nu_2, \nu_3; x, y)=\sum_{n_1, n_2, n_3}f_{222}(n_1, n_2, n_3;x, y) J(\nu_1-n_1, \nu_2-n_2, \nu_3-n_3; x, y),
\end{align}
The first line captures the decomposition of the three power spectra with the sum over $m_i$, coupled to their respective frequencies $\nu_i$, i.e. sourced from the decomposition Eq. \eqref{eq:fftlog_p11_decomp}. Additionally present is the sum over the kernels $F_2$ written in terms of polynomials of $q^{2 n_1}, |\k_1-\q|^{2n_2}, |\k_2+\q|^{2n_3}$, expressed by the sum over $n_1, n_2, n_3$ as well as the rational coefficient $f_{222}$ \cite{Simonovi__2018}.

The decomposition of momentum field kernels $M_n^z$ into a polynomial is complicated as they involve a specific line of sight direction. For example, the decomposition of $M^z_{2}(\k_1+\q,\k_2-\q)$ is written in terms of $q$, $| \k_1-\q|$, $|\k_2 + \q|$, but also $(\k_1-\q) \cdot \hat{\z}$, $(\k_2+\q) \cdot \hat{\z}$. Expanding these dot products, we can pull out dependencies on $\k_{1,2} \cdot \hat{\z} = k_{1,2} \mu_{1, 2}$, but $\hat{\q} \cdot \hat{\z}$ remains inside the integral. To deal with $\hat{\q} \cdot \hat{\z}$, we use the fixed vectors $\k_1, \k_2$ as a basis, expanding $\hat{\q} \cdot \hat{\z}$ in terms of $\mu_1, \mu_2$, and $\nu_{12}\equiv \hat{\k}_1 \cdot \hat{\k}_2$. Since our bispectrum involves only a single power of the momentum field, we have a single power of $\hat{\q} \cdot \hat{\z}$, so we only require a decomposition like:

\begin{align}\label{eq:qdotz_first_decomp}
    \hat{\q} \cdot \hat{\z} = \frac{\mu_1 - \mu_2 \nu_{12}}{1-\nu_{12}^2} \hat{\q} \cdot \hat{\k}_1 + \frac{\mu_2 - \mu_1 \nu_{12}}{1-\nu_{12}^2} \hat{\q} \cdot \hat{\k}_2\,,
\end{align}
as shown by \cite{Philcox:2022frc}. Terms like $\hat{\q} \cdot \hat{\k}_1$ can be rewritten as a polynomial in terms of the desired quantities via $|\k_1-\q|^2 = k_1^2+q^2-2 k_1 q \;\hat{\k}_1\cdot\hat{\q}$. This decomposition of $\hat{\q} \cdot \hat{\z}$ enables us to express the integrands components of an angular basis, which looks like
\begin{align}\label{eq:fftlog_ang_basis}
    &B_{222}(\k_1, \k_2, \k_3) 
    \Rightarrow B_{222}(k_1, k_2, k_3, \mu_1, \mu_2) = \mu B_{222, \mu}(k_1, k_2, k_3) + \chi B_{222, \chi}(k_1, k_2, k_3),
\end{align}
where $\mu=\mu_1$ and $\chi\equiv \sqrt{1-\mu^2}\cos(\phi)$, such that the three $\mu_i$ angles are given by:
\begin{align}\label{eq:mu_defns}
    \mu_1=\mu, \qquad \mu_2 = \mu \nu_{12} - \chi \sqrt{1-\nu_{12}^2}, \qquad \mu_3 = \frac{-k_1 \mu_1-k_2\mu_2}{k_3}.
\end{align}

This $\mu, \chi$ decomposition is simply a redefinition of the angular basis in Eq. \eqref{eq:qdotz_first_decomp}. We also note that we have principally gone from the 6 degrees of freedom in $B_{222}(\k_1, \k_2, \k_3)$ to 5 via the symmetry of rotation about the line of sight direction. Now, these $B_{222, \mu}$ and $B_{222, \chi}$ pieces can be computed following the usual formalism Eq. \eqref{eq:fftlog_b222_explicit}, decomposing the integrand into polynomials in the three vector magnitudes and evaluating the $J$ functions on a grid of $k_1, x, y$, equivalently different triangle configurations $k_1, k_2, k_3$. 

\subsection{Comparison against Numerical Integration}
As a precision check of FFTLog, we make a comparison against brute force integration for a specific choice of triangle geometry (i.e. $k_2^2/k_1^2, k_3^2/k_1^2, \mu_1, \mu_2$) over a range of $k_1$, here applying the \texttt{vegas} algorithm \cite{1978JCoPh..27..192L, Lepage_2021}. The implementation to perform these 1-loop numerical integrals was built from the work by Ref.~\cite{deBelsunceSullivanMcDonald2025}. Here, we set the FFTLog bias $\nu=-1.6$, such that the leading IR limits look like $\int_{|\q| \ll k} P(q)/q^2 \sim\int dq \;q^\nu$. In the limit $q\rightarrow0$, this integral is non-convergent for $q < -1$. Given the nature of FFTLog, these leading IR limits are then implicitly subtracted from the results. As such, to match each loop term, those leading IR limits must be explicitly subtracted from the numerical integration. The equivalence principle asserts that all IR limits must cancel once the 1-loop contributions are summed together \cite{Kehagias:2013yd, Peloso:2013zw, Creminelli:2013mca, Baldauf:2015xfa,Blas:2013bpa,Blas:2015qsi}, which we have checked explicitly. For choices of FFTLog parameters described below, we find a percent-level match, which can be further improved by 
increasing the precision settings of the FFTLog routine. 

\begin{figure}[H]
    \centering
    \includegraphics[width=\linewidth]{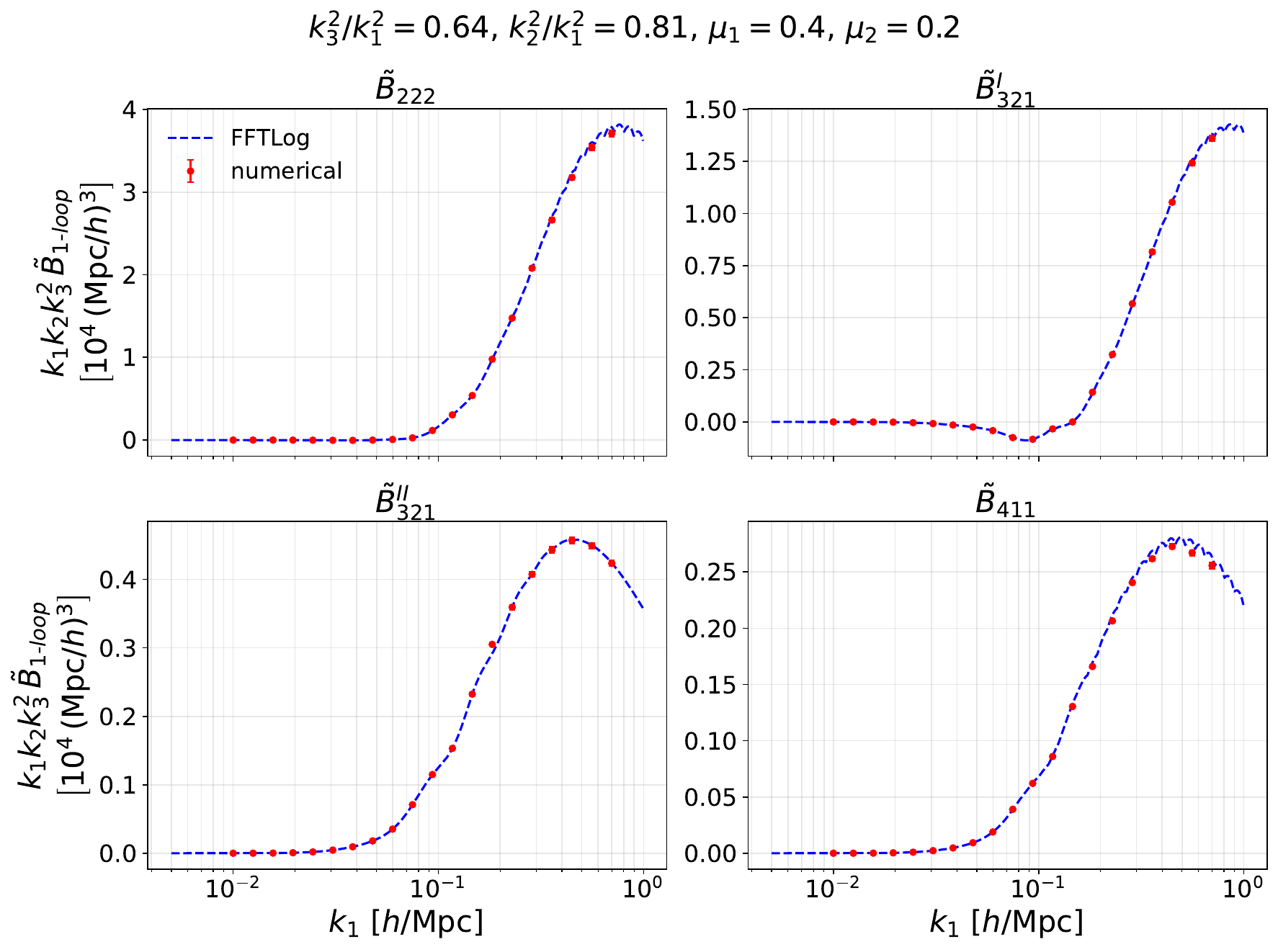}
    \caption{Comparison of FFTLog \cite{Simonovi__2018} (blue) against brute-force numerical integration via the \texttt{vegas} algorithm \cite{1978JCoPh..27..192L, Lepage_2021} (red) for each of the 1-loop corrections in Eq. \eqref{eq:bispec_loops}, with leading IR limits subtracted. Error bars from numerical integration are of order few $\%$, therefore not easily visible. Visual agreement across most of the range of $k_1$ is evident for each loop correction, with slight ringing artifacts appearing towards higher $k$. Shown is the tilde bispectrum, using the notation to remove the factor of $i f\HH$ from the momentum field. We have also multiplied by an extra factor of $k_3$ to account for the $1/k$ in the momentum field.
    }
    \label{fig:fftlog_vs_numint_individual}
\end{figure}

The check of FFTLog against numerical integration validates our calculation of the 1-loop corrections at the few percent level, particularly in the range where we will ultimately apply our theory, $k\lesssim0.4\:\iM$.
The FFTLog computational cost is dictated by the number of $N_{\rm samp}$ points used to decompose the power spectrum in Eq. \eqref{eq:fftlog_p11_decomp}, as well as the resolution of the grid $k_1, x, y$. Particularly, the presence of three power spectra in $B_{222}$ requires a sum over $N_{\rm samp}^3$ terms, taking much of the overall run-time across loop corrections. 
To run FFTLog, we have adopted $N_\text{samp}=64$ for $B_{222}$, a grid of $x^2, y^2$ with $40$ points from $0$ to $1$, and evaluated at $64$ logarithmically spaced values of $k_1$ from $0.001 \: \iM$ to $1 \: \iM$. The power spectrum decomposition is done in the range $[10^{-5}, 180]\: \iM$. While sufficient for the work in this paper, these choices leave much room for improvement of precision. In addition to better resolving the power spectrum in decomposition, a finer grid of triangle configurations reduces uncertainties in interpolation, which are then propagated to the finite-bin integration discussed in Section~\ref{sec:binning}. Nonetheless, the precision demonstrated here is sufficient for comparisons made against simulation in this paper. 
\begin{figure}[H]
    \centering
    \includegraphics[width=\linewidth]{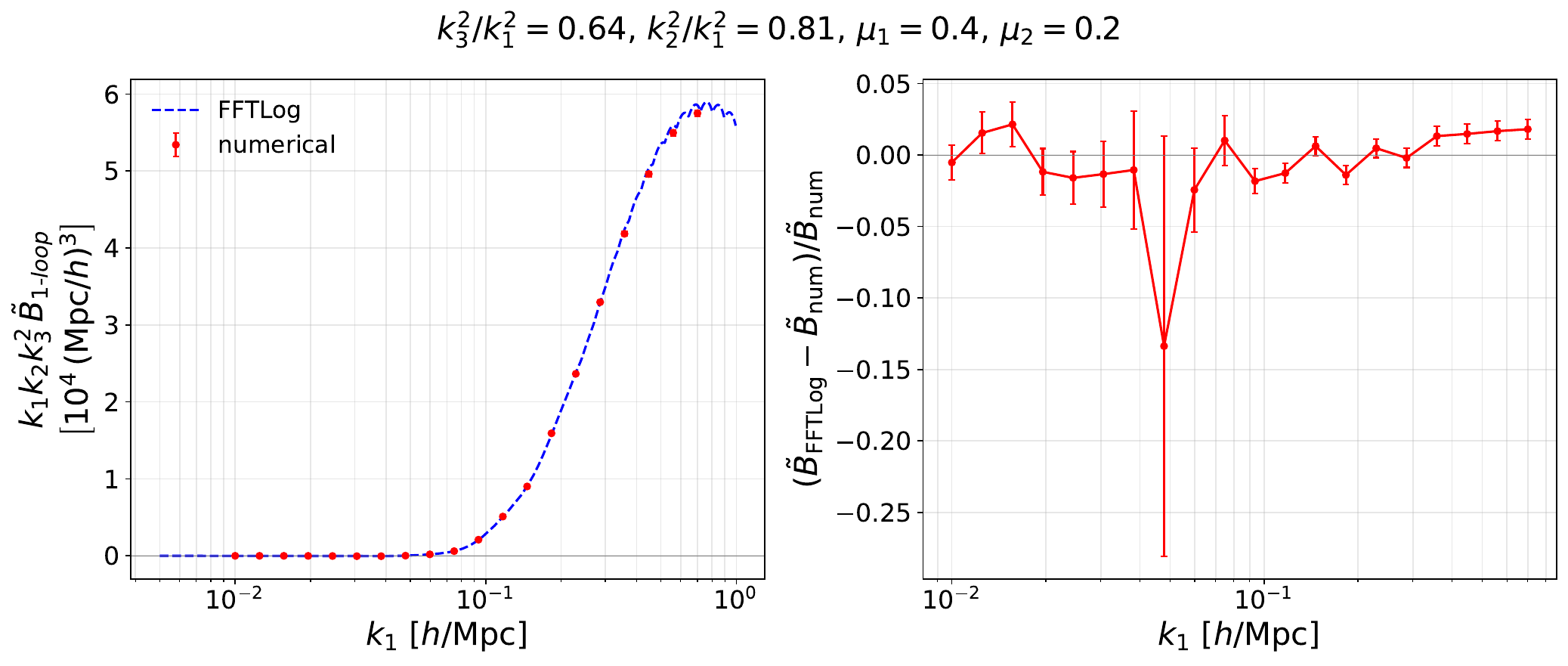}
    \caption{Comparison of FFTLog \cite{Simonovi__2018} (blue) against numerical integration (red) for the entire 1-loop correction, i.e. the combination of panels in Fig.~\ref{fig:fftlog_vs_numint_individual}. A fractional comparison validates FFTLog at the few percent level in the regime of validity determined via a comparison to simulation, sufficient for work in this paper. Error bars are from numerical integration with \texttt{vegas}.%
    }
    \label{fig:fftlog_vs_numint_summed}
\end{figure}

\section{UV/IR Limits and Counterterms \label{sec:uv}}
In this Section we explicitly include the EFT counterterms, which remove the 
UV sensitivity of the SPT loop integrals. The dark matter counterterms
appear in the effective dark matter stress-tensor $\tau_{ij}$ in the 
r.h.s. of the Euler equation.
They consist of all possible 
operators consistent with symmetries
of the problem (rotation invariance and the equivalence principle).
The free parameters in front of these 
operators are the Wilson coefficients,
which absorb the UV sensitivity,
i.e. the mistake one does by extrapolating
the SPT loops to arbitrarily small 
scales. 

\subsection{Counterterms - Power Spectrum}
In SPT, when solving the cosmological fluid equations, one makes the assumption that the contribution from the stress energy tensor $\partial_j \tau^{ij}$ is zero. In reality, this is flawed: when integrating out small scales, one sees that these scales couple to the large-scale predictions, so even to make predictions on large scales the stress energy tensor cannot be ignored. EFT is a way of formalizing this: contributions from the stress-energy tensor are inherited by the density and momentum fields via the Euler Equation \eqref{eq:exact}, and those contributions are precisely the counterterms that serve to renormalize UV divergences appearing in SPT. Writing the stress-energy tensor to leading order in derivatives and density fields, we have \cite{Baldauf:2014qfa} (but slightly changing notation):
\begin{align}
    \tau^{\text{vis}}_\theta\big|_{\text{LO}} \equiv \partial_i \tau^i_{\text{vis}}\big|_{\text{LO}} = -c_s^2\partial_i\partial^i\delta_{1},
\end{align}
where the $\text{vis}$ superscript indicates that this is the contribution from the viscosity of the fluid, as opposed to the stochastic contribution which is irrelevant for pure matter fields~\cite{Pajer:2013jj,Baldauf:2015zga,Ivanov:2026nlf}. By inspecting Equation \eqref{eq:exact}, we see that at the same order, the density field inherits a counterterm correction like
\begin{align}
    \delta^{\text{ctr}}(\k, \tau)\big|_{LO}= -\gamma(\tau) k^2\delta_{1}(\k, \tau),
\end{align}
 where $\gamma$
 is the effective sound speed of
 dark matter. 
This counterterm enables the renormalization of the UV divergent $P_{13}$-type integral in $P_{\delta\delta}$, shown explicitly in Eq. \eqref{eq:power_spec_ctrs}. Analogously, necessary to renormalize the $P_{\delta\pi}$ and $P_{\pi \pi}$ power spectra is the counterterm correction for the momentum field, which we can easily relate to the density field counterterm via the continuity equation:
\be 
\begin{split}\label{eq:m1ctr_from_cont_0}
    i k_i [\pi^{\text{ctr}}(\k, \tau)]^i_{\text{LO}} &= 
    -\partial_\tau \delta^{\text{ctr}}(\k, \tau) \big|_{LO}
    = \partial_\tau[\gamma(\tau) \delta_{1}(\k, \tau)]\\
    &= k^2 \left[ f\HH \frac{d \gamma}{d\ln D}\delta_1(\k, \tau)+\gamma \partial_\tau \delta_1(\k, \tau) \right]\\
    \Rightarrow \quad [\pi^{\text{ctr}}(\k, \tau)]^i_{\text{LO}} &= -i k_i \left[\frac{d \gamma}{d\ln D}+\gamma \right]f\HH \delta_1(\k, \tau)\,,
\end{split}
\ee 
where $D(\tau)$ is the growth factor,
and we have used the linear result $\partial_\tau \delta_1(\k, \tau)=f \H \delta_1(\k, \tau)$~\cite{Baldauf:2014qfa}. From this structure we can write the effective kernels contributed to the density and momentum fields by the counterterms as
\begin{align}
    F_1^{\text{ctr}}(\k) &= -k^2 \gamma\\
    [M_1^{\text{ctr}}]^i(\k) &= -i k^i 
   \left( \gamma+\frac{d\gamma}{d\ln D}\right)f\HH \,.
\end{align}
These lead to the following power spectrum corrections (for $i=z$):
\begin{align}\label{eq:power_spec_ctrs}
    P_{\delta \delta}^{\text{ctr}}(k, \tau)\big|_{LO} &= -2 k^2 \gamma(\tau) P_{11}(k, \tau)\notag\\
    \tilde{P}_{\delta \pi_z}^{\text{ctr}}(\k, \tau)\big|_{LO} &=- k_z 
    \left[2\gamma(\tau)+\frac{d\gamma}{d\ln D}\right] P_{11}(k, \tau) \notag\\
    \tilde{P}_{\pi_z\pi_z}^{\text{ctr}}(\k, \tau)\big|_{LO} &= -2 \frac{k^2_z}{k^2} \left[\gamma(\tau)+\frac{d\gamma}{d\ln D} \right]P_{11}(k, \tau)\,.
\end{align}
Note that these are the 
most general expressions that do not make assumptions about the time-dependence of $\gamma$.
However, some estimates for the time-dependence of $\gamma$ are available. 
Specifically, this counterterm
has two parts: the ``infinite'' one 
that cancels the spurious UV sensitivity
of the SPT loops, and the ``finite'' part
that captures the physical backreaction
from small-scales~\cite{Ivanov:2026nlf}. 
Using the ansatz $\gamma(\tau) = \gamma_0 D^{m+1}(\tau)$, these correspond to $m=1$
and $m=5/3$, respectively, where in the last case we also approximated the linear power spectrum as a power law with the slope $n=-3/2$. In that case we can substitute in the above equations:
\be \label{eq:m1ctr_from_cont}
\frac{d\gamma}{d\ln D}\to \gamma( m+1)\,,
\ee 
producing the coefficients $\gamma(3+m)$
and $\gamma(2+m)$ in $P^{\rm ctr}_{\delta \pi_z}$
and $P^{\rm ctr}_{\pi_z\pi_z}$,
respectively.

\subsection{Power Spectrum Limits: UV Renormalization and IR Effects}

Here we compute directly the UV limits, taken as the leading behavior inside the integral under $q \gg k$, of the various 1-loop power spectra corrections derived in Section~\ref{sec:pk}. 
This will serve as a demonstration of the above counterterm expressions correctly renormalizing divergences. We focus on the case $i=z$ in what follows. We begin with the matter-matter and matter-momentum power spectra:

\begin{align}
    P_{\delta \delta}^{\text{1-loop}}\bigg|_{q \gg k}
&=
\underbrace{
\frac{9}{98}k^4 \int_\q \frac{P_{11}^2(q)}{q^4}}_{(22)}
-
\underbrace{
\frac{61}{315} k^{2}
P_{11}(k)\int_\q \frac{P_{11}(q)}{q^2}}_{(31)},\\
\tilde{P}_{\delta \pi}^{\text{1-loop}}\bigg|_{q \gg k}&=\underbrace{\frac{9}{49}\mu k^3 \int_\q \frac{P_{11}^2(q)}{q^4}}_{(22)} - \underbrace{\frac{122}{315} \mu k P_{11}(k) \int_{\q} \frac{P_{11}(q)}{q^2}}_{(31)},
\end{align}
where we have indicated the UV limits of the $P_{22}$-like and $P_{31}$-like corrections separately. Inspecting the structure of the $P_{31}$ divergences, we can see that they are exactly canceled by the respective counterterm expressions in Eqs. \eqref{eq:power_spec_ctrs}. By directly matching the expressions, we recover the standard dark matter sound speed renormalization condition~\cite{Baldauf:2014qfa}: 
\be 
\gamma+\frac{61}{630} \int_{\q} \frac{P_{11}(q)}{q^2}=\gamma_{\rm fin}\,, 
\ee 
suggesting the power law scaling $m=1$ for $\gamma_{\rm inf}$. 
The $P_{22}$ terms above
are formally renormalized
by the stochastic stress-tensor
components, but these contributions are beyond the one-loop order in EFT power counting~\cite{Ivanov:2026nlf} so we ignore them
in this work. 

We also compute the IR limits of the matter power spectrum and matter-momentum cross-spectrum, i.e. take $q \ll k$. The leading IR limits are known to cancel between $P_{22}$ and $P_{13}$
as dictated by the equivalence principle. We continue with the momentum-momentum power spectra, but now present both the UV as well as the non-canceling IR limits:
\begin{align}
    \tilde{P}_{\pi \pi}^{\text{1-loop}}\bigg|_{q \gg k} &= \underbrace{\left[\frac{2}{15}+ \mu^2 \frac{172}{735} \right] k^2 \int_q \frac{P_{11}^2(q)}{q^4}}_{(22)} - \underbrace{\frac{61}{105} \mu^2 P_{11}(k)\int_q \frac{P_{11}(q)}{q^2}}_{(31)},\notag\\
    \tilde{P}_{\pi \pi}^{\text{1-loop}}\bigg|_{q \ll k} &= \underbrace{\frac{1}{3}P_{11}(k) \int_q \frac{P_{11}(q)}{q^2}}_{(22)}.
    \label{eq:IRUVpipi}
\end{align}
We see again that the $P_{31}$ UV divergence is canceled with the respective counterterm expression in Equation \eqref{eq:power_spec_ctrs} upon choosing $m=1$. However, unlike the previous two power spectra, the leading IR limits do not cancel. This could be understood from the 
long-mode construction of~\cite{Baldauf:2015xfa,Ivanov:2026nlf}.
To leading order in
$q/k$, a mode $\q$ with $q \ll k$ 
sources a uniform bulk velocity $\v_L$ and the displacement $\bm{\Psi}_L = \int d\tau\, \v_L$. 
Its action on the short-scale fields 
can be understood as a boost under which the fields
transform as
\begin{align}\label{eq:ir_boost}
    \delta(\k) &\to e^{-i\k\cdot\bm{\Psi}_L}\,\delta(\k)\,, \qquad
    \theta(\k) \to e^{-i\k\cdot\bm{\Psi}_L}\,\theta(\k)\,, \notag\\
    \pi_i(\k) &\to e^{-i\k\cdot\bm{\Psi}_L}\left[\pi_i(\k) + v_{L,i}\,\delta(\k)\right]\,,
    \qquad (k \neq 0)\,,
\end{align}
where the additional $v_{L,i}\,\delta(\k)$ term arises because
of the Galilean transformation 
$\v \to \v(\q_L=\x - \bm{\Psi}_L) + \v_L$, where
$\q_L$ here is the Lagrangian coordinate. The momentum $\pi_i = (1+\delta)v_i$,
unlike the density or the velocity divergence, is not invariant under Galilean
boosts \cite{Mercolli_2014}. In any equal-time two-point function the phases
combine to $e^{-i(\k_1+\k_2)\cdot\bm{\Psi}_L} = 1$ identically, producing
the known 
equivalence-principle cancellation. 
What remains after the phases cancel is the
$v_{L,i}\,\delta$ term of Eq.~\eqref{eq:ir_boost}. In $P_{\delta\pi}$ it
appears linearly and averages to zero, $\langle v_{L,i}\rangle = 0$, so the
cross power spectrum is IR-safe. In $P_{\pi\pi}$ it appears quadratically and
contributes
\begin{align}
    \langle v_{L,i}\, v_{L,j}\rangle\, P_{11}(k)
    = \frac{\delta_{ij}}{3}\,\sigma_v^2\, P_{11}(k)\,, \qquad
    \sigma_v^2 \equiv (f\HH)^2 \int_\q \frac{P_{11}(q)}{q^2}\,,
\end{align}
which is exactly the $(22)$ IR limit in Eq.~\eqref{eq:IRUVpipi}.

The one-loop corrections (specifically $P_{22}$) to the momentum-momentum power spectrum introduce a transverse contribution with $\mu = 0$, whereas the term from linear theory scales exactly as $\mu^2$. This is critical for the case of the kSZ field, where the projection along the line of sight direction restricts to 2-d Fourier modes, i.e. $\mu=0$. This implies that any longitudinal contribution, i.e. that with $\partial_i \pi^i \neq 0$, does not appear in the kSZ field. The only way to source transverse contributions is through the product $\delta \v$ appearing in the momentum field, which is why any correlation function must involve the kSZ field at minimum second order. In terms of the existing literature, this is known as the $P_{q_{\perp} q_{\perp}}$ contribution \citep{Vishniac:1987_kSZ}.

In evaluating the one-loop corrections to the power spectrum via the FFTLog formalism, we make a choice of $\nu=-1.6$ (see Equation \eqref{eq:fftlog_p11_decomp}). Although $\nu=-1.6$ presents the most consistent match against brute force numerical integration, it lacks the presence of the leading IR contribution. In most cases, this isn't an issue, since the leading IR contributions cancel, but presents a problem for $P_{\pi_z \pi_z}$, as per the above discussion. As such, this non-canceling IR limit must be added back in by hand, requiring a numerical integration of $\int_q P(q)/q^2$. Such an integration is cutoff-dependent, i.e. depends on the maximum $q$ to which the integration is performed, and therefore the result cannot be trusted in the theoretical prediction. 
A principled way to resolve this is to
perform the integration
only over a specified 
IR domain determined by an IR separation scale $k_{\rm IR}$ and treat the residual dependence on the latter as theoretical error similarly
to how this is done in the case of IR resummation for dark matter~\cite{Blas:2016sfa}.
In our work, however, we follow a more straightforward approach: we do not include the monopole $P_{\pi_z \pi_z}^{\ell=0}$, which contains this residual IR contribution, in fitting the counterterm.

\subsection{Counterterms - Bispectrum}
In this section we present the counterterms necessary to renormalize UV divergences in the 1-loop bispectrum, leaving the majority of details
to Appendix \ref{app:M_n_ctr}. 
Our derivation follows Ref.~\cite{Baldauf:2014qfa}. However, 
we find that the contributions to the bispectrum counterterms 
from the time-integrals over the effective sound speed
identically reduce to 
the usual sound speed contribution multiplying 
$k^2\delta$. 
This reduction is required
by the equivalence principle
and happens irrespectively
of any assumptions about the 
time-dependence of $\gamma$.
While this result is implicitly present
in~\cite{Baldauf:2014qfa,Chen:2026usz,Ivanov:2026zos},
in our context
it changes the counting 
of free parameters in the bispectrum counterterm. 

To renormalize UV divergences appearing in the 1-loop bispectrum, it is necessary to compute the next order of counterterm corrections \cite{Baldauf:2014qfa, Angulo:2014tfa}. Intuitively, this is because the tree-level bispectrum is $\mathcal{O}(\delta^4)$ and the one-loop corrections scale like $\mathcal{O}(\delta^6)$, each two orders higher than the equivalent tree-level/1-loop corrections for the power spectrum. Schematically, this will resemble two contributions --- $B_{c21}$, with the first order counterterm, and $B_{c11}$, with the second order counterterm: 
\begin{align}
    B_{\text{mm}\pi}^{\text{ctr}}(\k_1, \k_2, \k_3; \tau) &= B_{c21}(\k_1, \k_2, \k_3; \tau) + B_{c11}(\k_1, \k_2, \k_3; \tau).
\end{align}
These second order counterterm kernels can be written as all second order derivative corrections~\cite{Baldauf:2015aha,Chen:2026usz,Ivanov:2026zos}. We choose to the following basis of these corrections:
\be 
F_2^{\text{ctr}}(\q_1,\q_2)
= -\sum_{i=1}^{3} \tilde{\epsilon}_i\, 
E_i(\q_1,\q_2)
       - \gamma k^2 F_2^{\rm SPT}(\q_1,\q_2)\,,
\ee 
where $F^{\rm SPT}_2$ is the SPT density kernel~\cite{Bernardeau:2001qr} and 
we have introduced the shape functions
\be 
 E_1=k^2, \quad 
E_2=k^2\Big(\frac{(\q_1\cdot\q_2)^2}{q_1^2q_2^2}-\frac13\Big), \quad 
 E_3=-\frac{k^2}{6}+
 \frac{(\q_1\cdot\q_2)}{2}\Big(\frac{\k\cdot\q_1}{q_1^2}+\frac{\k\cdot\q_2}{q_2^2}\Big)\,.
 \label{eq:E123}
\ee
 As detailed in Appendix \ref{app:M_n_ctr}, $M_2^{\text{ctr}}$ contains novel transverse structures inherited from
the 
sound speed and 
the unique transverse operator in the stress-energy tensor at the second order, yielding:
\begin{align}\label{eq:M_2_ctr}
  & M_2^{i, \text{ctr}}(\q_1, \q_2; \tau) =\frac{1}{D^2} \frac{i k^i}{k^2} \partial_\tau [D^2 F_2^{\text{ctr}}(\q_1, \q_2;\tau) ]\notag \\
    &\qquad - i\frac{f\HH}{2}\frac{q_2^i (\q_{12}\cdot \q_1) - q_1^i(\q_{12}\cdot \q_2)}{q_{12}^2}\left[ c_\omega^2 \left[ (\q_1 \cdot \q_2) \bigg( \frac{1}{q_1^2} - \frac{1}{q_2^2}\bigg) \right] + \gamma\left[ \frac{q_1^2}{q_2^2} - \frac{q_2^2}{q_1^2}\right]\right]\,,
\end{align}
with the longitudinal contribution in the first line with $\partial_\tau$ and the novel transverse contribution in the second line. With these kernels, we can write the overall counterterm contribution to the bispectrum, now explicitly including time-dependence:
\begin{align}\label{eq:bispec_ctr_structure}
    B_{c21}(\k_1, \k_2, \k_3; \tau)&= 2F_2(\k_3, \k_1)
     F_1(\k_1)\,
   P_{11}(k_3; \tau)\, P_{11}(k_1; \tau)
   M_1^{\text{ctr}}(-\k_3) + \{\k_1 \leftrightarrow \k_2\} \notag \\
   & \quad + \, 2M_2(\k_1,\k_2)\,
    F_1(\k_1)\,
   P_{11}(k_1; \tau)\, P_{11}(k_2; \tau)
   F_1^{\text{ctr}}(\k_2) + \{\k_1 \leftrightarrow \k_2\}\notag \\
   & \quad + \, 2F_2(\k_1,\k_3)\,
    M_1(-\k_3)\,
   P_{11}(k_1; \tau)\, P_{11}(k_3; \tau)
   F_1^{\text{ctr}}(\k_1) + \{\k_1 \leftrightarrow \k_2\}\\
	 B_{c11}(\k_1, \k_2, \k_3; \tau) &= \, 2 F_1(\k_1)\, F_1(\k_2)\,
   P_{11}(k_1; \tau)\, P_{11}(k_2; \tau)
   M_2^{\text{ctr}}(\k_1, \k_2; \tau)\notag\\
   &\quad+ \,2 F_1(\k_1)\, M_1(-\k_3)\,
   P_{11}(k_1; \tau)\, P_{11}(k_3; \tau)
F_2^{\text{ctr}}(\k_1, \k_3; \tau)+ \{\k_1 \leftrightarrow \k_2\},
\end{align}
where the corresponding counterterms to renormalize each of the $B_{321}^{II}$ and $B_{411}$ loops are apparent. We explicitly verify that the UV limits of $B_{321}^{II}$, $B_{411}$ are exactly canceled by these counterterm expressions above in Appendix \ref{app:bispec_uv_cancel}.

Without an assumption of the time-dependence of counterterm coefficients, there are 9 parameters: the sound speed and its time
derivative $\gamma, \frac{d\gamma}{d\ln D}$, three longitudinal coefficients for the second order counterterms and their time-derivatives $\{\tilde{\epsilon}_1, \tilde{\epsilon}_2, \tilde{\epsilon}_3 \},\frac{d}{d\ln D}\{\tilde{\epsilon}_1, \tilde{\epsilon}_2, \tilde{\epsilon}_3 \}$, and the transverse second order counterterm for the momentum field $c_\omega^2$. The UV-motivated ansatz~\cite{Baldauf:2014qfa,Steele:2020tak,Baldauf:2021zlt} provides a phenomenological model for the time-dependence of counterterms, which can reduce this set of parameters to two. 
The first is the first order counterterm coefficient $\gamma$. Matching the UV limit gives $\gamma=-\frac{61}{630}\sigma_d^2$ where we have defined $\sigma^2_d \equiv \int_\q P_{11}(q)/q^2$,
which allows us to formally
replace $\sigma_d^2 \to \sigma_d^2 -\frac{630}{61}\gamma $
in the expressions for the 
UV divergences. 
The second parameter, $m$, controls the power law scaling of the counterterms, allowing us to replace all time 
derivatives as in Eq. \eqref{eq:m1ctr_from_cont}. Under this phenomenological ansatz, the 9 counterterm coefficients can all be expressed in terms of $\gamma$ and $m$. Albeit a manifestly approximate scheme,\footnote{A similar scheme was previously applied to the matter bispectrum. It leads to a good match at the level of the summary statistics~\cite{Baldauf:2014qfa}, but ultimately fails at the field level, which allows one to measure all the bispectrum counterterms 
independently with high precision~\cite{Steele:2020tak}.} it serves as a useful way to correct the UV divergences without introducing many counterterms to be fit. We analyze its effectiveness in Section~\ref{sec:sims}. 

In Appendix \ref{app:M_n_ctr}, we demonstrate that $\vec{M}_2^{\text{ctr}}$ acquires a transverse component (so that $[\vec\partial \times \vec{M}_2^{\text{ctr}}]\neq 0$), which is necessary to renormalize the UV divergence in $B_{411}$. In Section~\ref{sec:binning} below, to check our theoretical bispectrum expression against measurements from a simulation, we compute the dipole over the angle $\mu_3$. As a result, the contribution from these transverse counterterms is zeroed, and the comparison of the bispectrum dipole over $\mu_3$ has 8 principally free coefficients. In reality, the actual observable is the 2-d kSZ field, corresponding to the momentum field integrated over the line of sight. The 2-d projected field can be thought of as a subset of the 3-d momentum field keeping only Fourier modes with $\mu_3=0$, i.e. removing any longitudinal contributions and preserving only the transverse ones. In Section~\ref{sec:sims} we include a comparison between simulation and theory for the bispectrum involving the projected momentum field, requiring only the transverse counterterms. While this bispectrum is quite noisy relative to full 3-d momentum field, it nonetheless demonstrates the importance of the transverse counterterms for the modeling of the realistic kSZ field.

Before closing, let us note that this section neglects the UV divergences of the $B_{222}$ and $B_{321}^I$ and the corresponding counterterms. The justification follows analogous to that of the $P_{22}$ UV divergence in the power spectrum, namely that $B_{222}$($B_{321}^{I}$) are renormalized by stochastic (mixed-stochastic) terms, 
whose effect is beyond the one-loop order,
which can be justified by a power-counting argument \cite{bakx2025oneloopgalaxybispectrumconsistent}. 

\section{Simulation Bispectrum Estimation and Theory Binning \label{sec:binning}}
Now that we have developed a framework to generate and efficiently evaluate theoretical predictions for the $\langle \delta \delta \pi_z\rangle$ bispectrum at the 1-loop order with counterterm corrections, we would like to make a comparison against simulation for a sense of the range of validity of this calculation. Given the finite and discrete nature of the simulation, we would like to average together each noisy realization of $B_{\delta \delta \pi_z}(\k_1, \k_2, \k_3)$ into a quantity comparable to theory. Similar to the power spectrum, we proceed by computing an angular multipole of the bispectrum in finitely sized bins.
The first part of this section describes this scheme for the simulation, followed by the analogous procedure to take angular multipoles of and bin the theoretical expressions.
\subsection{Simulation Bispectrum Measurement}
Starting with the simulation, we are interested in how to convert gridded density and momentum fields at a particular snapshot into a binned bispectrum multipole. We apply the Scoccimarro estimator formalism \cite{Scoccimarro:2015bla}, and present a brief overview of the scheme. Starting with the definition of a bispectrum multipole, we have: 
\begin{align}
    B_{\ell}^{mm\pi}(k_1, k_2, k_3) &\equiv \frac{2\ell+1}{2}\int_0^{2 \pi} d \phi \int_{-1}^1 d\mu_3 \;B^{mm\pi}(\k_1, \k_2, \k_3) \; \mathcal{L}_{\ell}(\mu_3) \notag \\
    &= \frac{2\ell+1}{2}\int_0^{2 \pi} d \phi \int_{-1}^1 d\mu_3 \;B^{mm\pi}(k_1, k_2, k_3, \mu_1[\mu], \mu_2[\mu, \phi]) \; \mathcal{L}_{\ell}(\mu_3),
\end{align}
where the definition of $\phi$ matches Eq. \eqref{eq:mu_defns}, and we have defined $\mathcal{L}_\ell(\mu)$ as the Legendre polynomial of order $\ell$. 
As only a singular angular power is present from the momentum field, we choose to compute only the dipole ($\ell=1$) moment. We also compute the dipole over $\mu_3$ to preserves the symmetry between $k_1$ and $k_2$.%
For the bispectrum with the projected momentum field, relevant to application to the kSZ, we have $\mu_3=0$. This means that a dipole must be taken over one of the angles $\mu_1$ or $\mu_2$, but since they are constrained by $k_1\mu_1+k_2\mu_2=0$, the choice becomes irrelevant.%

Computing the multipolar bispectrum for the remaining degrees of freedom $k_1, k_2, k_3$ involves averaging over the product $\delta(\q_1)\delta(\q_2)\pi_z(\q_3) \mathcal{L}_{\ell}(\hat{\z}\cdot\hat{\q}_3)$ for all $|\q_1|=k_1$, $|\q_2|=k_2$, and $|\q_3|=k_3$. In reality, fields are discrete grids, so we compute the bispectrum multipole over finitely-sized bins:
\begin{align}\label{eq:bispec_estimate_sum}
    B_{\ell}(k_1, k_2, k_3) &\equiv \frac{1}{V}\frac{2\ell+1}{2}\frac{1}{\mathcal{N}(k_1, k_2, k_3)} \sum_{\q_1+\q_2+\q_3=0} \delta_m(\q_1) \delta_m(\q_2) \pi_z(\q_3) \mathcal{L}_{\ell}(\hat{\z}\cdot\hat{\q_3})\notag\\ &\hspace{2.35in}\times\Theta_{k_1}(\q_1)\Theta_{k_2}(\q_2)\Theta_{k_3}(\q_3),
\end{align}
where the sum is taken over all wave-vectors $\q_1, \q_2, \q_3$ on the grid such that momentum conservation, i.e. $\q_1+\q_2+\q_3=0$ is satisfied. $V$ is the volume of the simulation, and the $\Theta$ functions enforce the bin restrictions under some width $\Delta k$ via
\begin{equation}
    \Theta_{k_1}(\q_1) \equiv \begin{cases}
        1 & \text{if } k_1-\frac{\Delta k}{2} \leq q_1 \leq k_1+\frac{\Delta k}{2},\\
        0 & \text{otherwise,}
    \end{cases}
\end{equation}
and $\mathcal{N}(k_1, k_2, k_3)$ counts the number of valid triangle configurations for a given $k_1, k_2, k_3$ to normalize the average, which we can write as 
\begin{align}
    \mathcal{N}(k_1, k_2, k_3) \equiv \sum_{\q_1+\q_2+\q_3=0}\Theta_{k_1}(\q_1)\Theta_{k_2}(\q_2)\Theta_{k_3}(\q_3),
\end{align}
 For an efficient implementation, we would like to rewrite the sum Eq. \eqref{eq:bispec_estimate_sum} into an expression in terms of simple manipulations of the fields - summing over individual Fourier modes for each $\k_1, \k_2, \k_3$ while simultaneously checking $\k_1+\k_2+\k_3=0$ would be prohibitively slow. We begin by taking the continuum limit:
\begin{align}\label{eq:bispec_cont}
    B_{\ell}(k_1, k_2, k_3) &\equiv \frac{2\ell+1}{2}\frac{1}{\mathcal{N}(k_1, k_2, k_3)}\int_{q_1,q_2,q_3} (2 \pi)^3 \delta_D^{(3)}(\q_{123}) \left[\delta(\q_1) \delta(\q_2)\pi_z(\q_3)\right]\notag\\& \hspace{2.27in} \times\Theta_{k_1}(\q_1)\Theta_{k_2}(\q_2)\Theta_{k_3}(\q_3) \mathcal{L}_{\ell}(\hat{\z} \cdot \hat{\q}_3)
\end{align}
The Scoccimarro estimator \cite{Scoccimarro:2015bla} rewrites this integral in terms of bandpassed fields, i.e. the real-space field restricted to Fourier modes in the bin of interest:
\begin{align}
    &I_{k_j}^{\ell}(\x) = \int_q e^{i \q \cdot \x} \;\Theta_{k_j}(\q) \delta(\q)\;\mathcal{L}_{\ell}(\hat{\z}\cdot \hat{\q})\text{, such that}\notag \\
    &B_{\ell}(k_1, k_2, k_3) = \frac{2\ell+1}{2}\frac{1}{\mathcal{N}(k_1, k_2, k_3)} \int d^3 \x \; I^{\ell=0}_{k_1}(\x)I^{\ell=0}_{k_2}(\x)I^{\ell=\ell}_{k_3}(\x).
\end{align}
It is straightforward to show that this definition holds by inserting the definitions $I^{\ell}_{k_j}$ into that of the final bispectrum estimator. In practice, we compute sets of bandpassed fields for each desired $k$ bins. For some $k_1, k_2, k_3$ at which we evaluate $B^\ell(k_1, k_2, k_3)$, this above integral can easily be performed as a sum over all spatial points.

\subsection{Binned Theoretical Bispectrum}
The FFTLog formalism described in Section~\ref{sec:fftlog} enables us to evaluate the 1-loop bispectrum for any valid triangle configuration. To match the theoretical predictions to the binned bispectrum multipole estimate from simulation, we must identically integrate over $\mathcal{L}_\ell(\mu_3)$ and finite $k$-bins. We begin with the continuum limit of the bispectrum estimator, Eq. \eqref{eq:bispec_cont}, and substitute the theoretical template evaluated by FFTLog:
\begin{align}
    B_{\ell}(k_1, k_2, k_3) &\equiv \frac{2\ell+1}{2\mathcal{N}(k_1, k_2, k_3)}\int_{q_1,q_2,q_3} (2 \pi)^3 \delta_D^{(3)}(\q_{123}) \left[\delta(\q_1) \delta(\q_2)\delta(\q_3)\Rightarrow B^{mm\pi}_{\text{theory}, \ell}(\q_1, \q_2, \q_3)\right] \notag\\& \hspace{1.8in} \times\Theta_{k_1}(\q_1)\Theta_{k_2}(\q_2)\Theta_{k_3}(\q_3) \mathcal{L}_{\ell}(\hat{\z} \cdot \hat{\q}_3)
\end{align}
To proceed, we take the integral over the delta function. Taking Eq. 91 by Ref.~\cite{bakx2025oneloopgalaxybispectrumconsistent} without the presence for Alcock-Paczynski distortions:
\begin{align}\label{eq:theory_binned}
   & B_{\ell}^{\text{theory, binned}}(k_1, k_2, k_3) = \frac{2\ell+1}{2\mathcal{N}(k_1, k_2, k_3)} \prod_{i=1}^3 \left[\int_{k_i-\Delta k/2}^{k_i+\Delta k/2} q_i dq_i \right] \notag\\
    &\qquad \qquad \times\int_0^{2 \pi} d \phi \int_{-1}^1 d\mu\; \mathcal{I}(q_1, q_2, q_3, \mu, \phi) B^{mm\pi}_{\text{theory}}(q_1, q_2, q_3, \mu_1[\mu], \mu_2[\mu, \phi]) \; \mathcal{L}_{\ell}(\mu_3),
\end{align}
 where $\mathcal{I}(q_1, q_2, q_3, \mu, \phi)$ is one if and only if $q_1, q_2, q_3$ can form a valid closed triangle and zero otherwise, serving to exclude invalid non-momentum conserving triangle configurations. In analog to the simulation bispectrum estimator, $\mathcal{N}(k_1, k_2, k_3)$ counts the number of valid triangles in a bin, but now in the continuum limit:
\begin{align}
    \mathcal{N}(k_1, k_2, k_3)=\prod_{i=1}^3\left[\int_{k_i-\Delta k/2}^{k_i+\Delta k/2} q_i dq_i \right] \int_0^{2 \pi} d \phi \int_{-1}^1 d\mu\; \mathcal{I}(q_1, q_2, q_3, \mu, \phi)
\end{align}

With the explicit FFTLog angular dependence Eq. \eqref{eq:fftlog_ang_basis}, the integrals over $\phi, \mu$ in Eq. \eqref{eq:theory_binned} can be analytically performed, reducing the dimensionality of numerical integration to the three integrals over finite width $k$-bins. To incorporate the counterterm corrections introduced in Section~\ref{sec:uv}, we follow this formalism identically, integrating over $\mathcal{L}_{\ell}(\mu_3)$ and the finite $k$-bin widths.

\section{Comparison between Theory and Simulation}
\label{sec:sims}

Having computed both the theoretical predictions for the power spectra and bispectra between Sections \ref{sec:pk}, \ref{sec:bispectrum}, and \ref{sec:uv}, as well as the machinery to estimate binned bispectrum multipoles from discrete simulation grids in Section \ref{sec:binning}, we continue with the comparison of theoretical models against simulation. We quantify the range of validity of a given theoretical model as $k_{\rm max}$, the largest $k$ scale at which theoretical predictions deviate from simulation measurements by $5\%$. While this is in principle an arbitrary choice, it better reflects the application to a real data analysis where measurements are made with noise. 
We begin by detailing the simulations used, followed by the comparisons of theoretical prediction against simulation for the power spectra, then the bispectrum with the 3-dimensional momentum field, and finally the kSZ-like bispectrum with the 2-dimensional projected momentum field.

\subsection{FLAMINGO Simulations}
Here, we detail the FLAMINGO simulations used \cite{Schaye_2023} and computation of discrete density and momentum grids.
For both the power spectra and 3-d momentum field bispectrum, we use the L5p6 dark matter only (DMO) simulation --- a $(5.6 \,\text{Gpc})^3$ N-body simulation with the D3A cosmology \cite{DES:2021wwk} and $5040^3$ dark matter particles (as well as $2800^3$ neutrino particles). We take the $z=0.5$ snapshot. The choice to neglect feedback mechanisms is consistent with the range of scales considered in this work. To estimate the 2-dimensional projected momentum field bispectrum, we use the L11p2 DMO simulation, identical except with side length $(11.2 \, \text{Gpc})^3$, i.e. 8 times the volume. To project the momentum field, we sum the 3-dimensional momentum field over the line of sight axis, corresponding to the snapshot geometry by \cite{Smith26PhRvD:ksz_bisp}. Here, the assumption is made that factors in the line of sight integral sourcing the kSZ (Equation \eqref{eq:ksz_integral_defn}) except $\hat{n}\cdot\bf{\pi}$ vary slowly with redshift. 

To compute the discrete grids, raw particle positions and momenta are deposited with the CIC scheme implemented in \texttt{Pylians} \cite{Pylians}. De-convolution by the CIC window is performed when grids are transformed into Fourier space, as required for power spectrum and bispectrum calculations. For the $(5.6 \: \rm Gpc)^3$ box used for power spectrum and 3-d momentum field bispectrum calculations, fields are natively deposited with $N_{\rm grid}=1024$, yielding a Nyquist frequency $k_{\text{Ny}}=0.844 \: \iM$, well beyond the range we expect our theoretical predictions to work to given the late time $z=0.5$. 

When computing the bispectrum for the 3-d momentum field, we use an effective $\text{N}_{\rm grid}=768$ for the Fourier transform to lower the required memory, imposing a Nyquist frequency $k_{\text{Ny}}=\pi \cdot768 / (3813\: \M)=0.633 \:\iM$. In turn, this means that we only trust triangles with $k_1+k_2+k_3< 2 \:k_{\text{Ny}}$ \cite{Sefusatti_2016}. On imposing a $k_{\rm max}$ for any single triangle side, this is most constraining for equilateral triangles: $k_{\rm max}^{\text{eq}} = 2/3 \: k_{\text{Ny}}=0.42\:\iM$. On computing the bispectrum, we use bin centers from $0.05\: \iM$ to $0.49\: \iM$ with a bin width $0.02\: \iM$, throwing out configurations past the Nyquist limit. Additionally, as the evaluation of the hypergeometric function in Eq. \eqref{eq:fftlog_J_defn} hits a singularity for flattened triangles, we do not consider triangle bins for which the bin-integration of theory encounters this flattened limit. However, this numerical issue can be corrected using specific expressions for 1-loop integrals for flattened triangles, as described by \cite{Philcox:2022frc, bakx2025oneloopgalaxybispectrumconsistent}. 

To compute the bispectrum for the 2-dimensional projected momentum field with the larger $(11.2 \; \rm Gpc)^3$ simulation, we adopt $N_{\rm grid}=2048$ to match the Nyquist frequency of the $(5.6 \; \rm Gpc)^3$ simulation box. Although computing bispectra with these larger grids requires much more memory, this is partially compensated by the fact that we only involve the 2-dimensional projected momentum field, of size $N_{\rm grid}^2$ as opposed to $N_{\rm grid}^3$. Due to this increased memory requirement, we choose not to use the $(11.2 \; \rm Gpc)^3$ box for the full 3-d momentum field bispectrum calculation. 
\subsection{Power Spectrum Fits}
\begin{figure}[t!]
    \centering
    \includegraphics[width=0.49\linewidth]{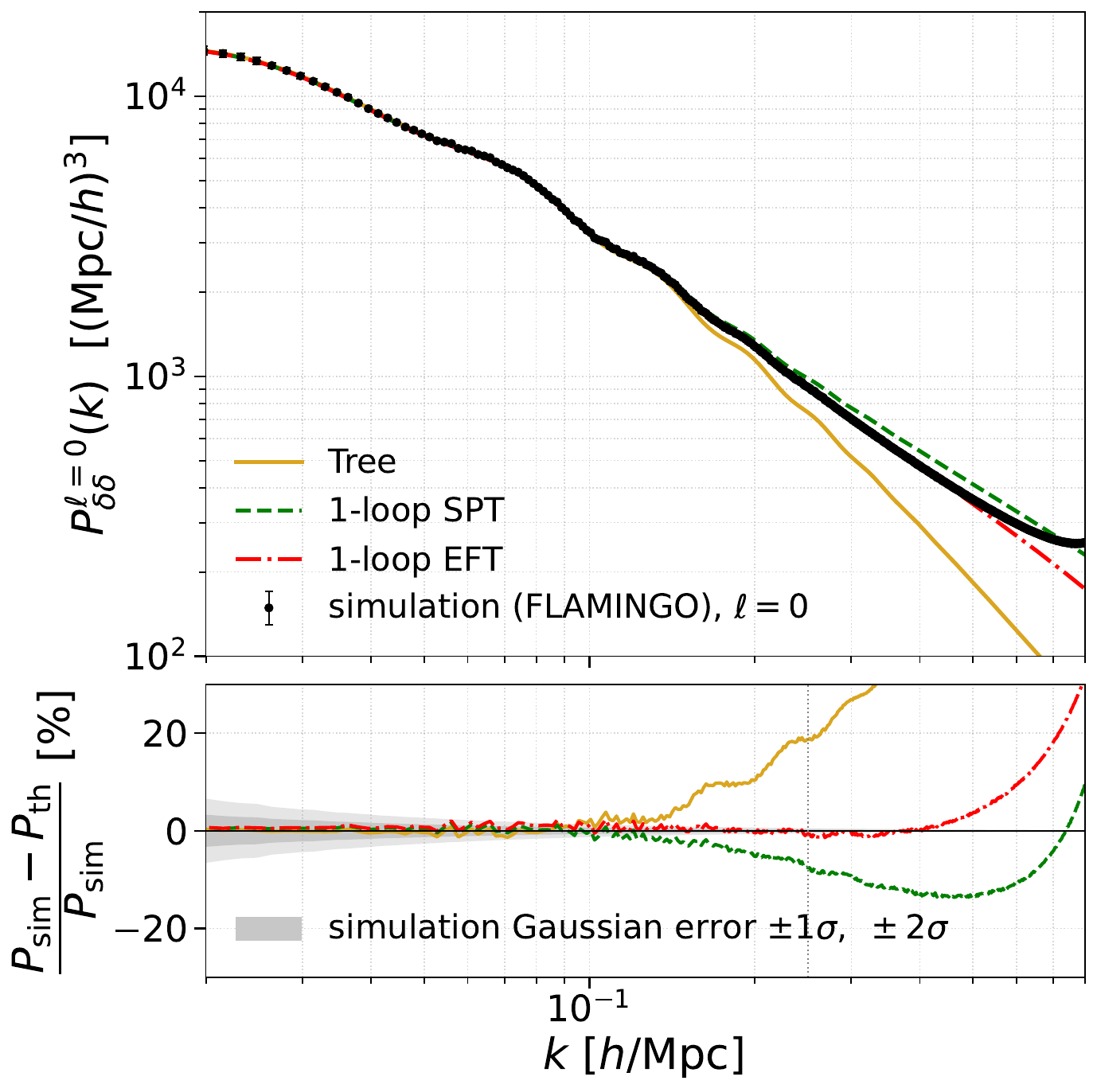}\hfill
    \includegraphics[width=0.49\linewidth]{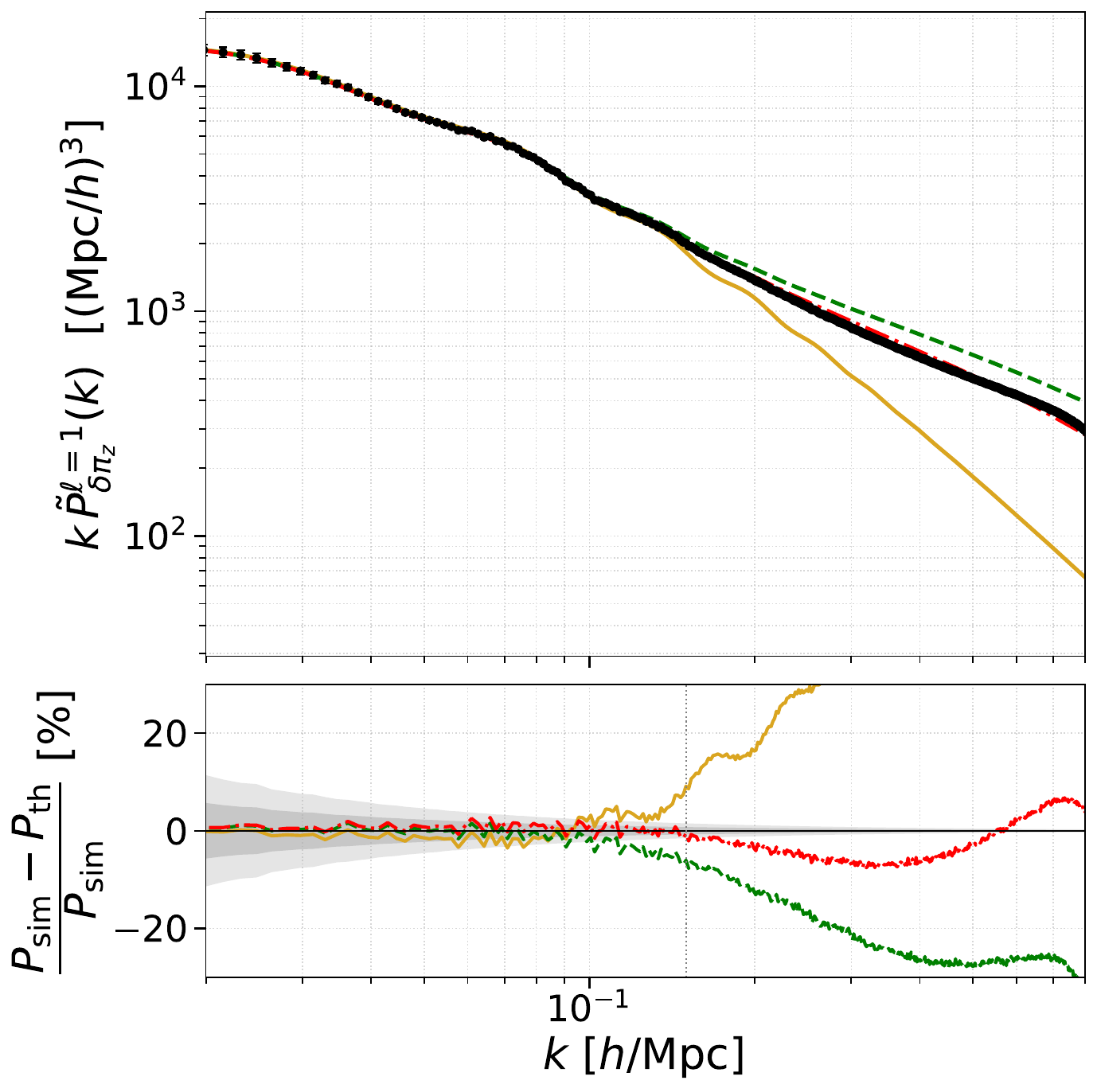}
    \caption{Comparison of tree-level, 1-loop SPT, and 1-loop EFT power spectra against FLAMINGO. \textit{Left:} $P_{\delta \delta}(k)$ monopole, with the counterterm coefficient fitted up to $k=0.25\:\iM$. \textit{Right:} $P_{\delta\pi_z}(k,\mu)$ dipole, with the counterterm coefficient fitted up to $k=0.15\:\iM$. Both demonstrate the expected improvement over tree-level (yellow) from adding the 1-loop corrections (green, 1-loop SPT), as well the fitted counterterm correction (red, 1-loop EFT). Roughly, as the momentum field is more non-linear \cite{Baldauf:2015aha}, we see that the range of agreement for $P_{\delta\delta}(k)$ is larger as compared to $P_{\delta \pi_z}$(k). Shown in shaded regions are the Gaussian error bars on simulation measurements. 
    }
    \label{fig:pk_mm_mpi}
\end{figure}

\begin{figure}[H]
    \centering
    \includegraphics[width=\linewidth]{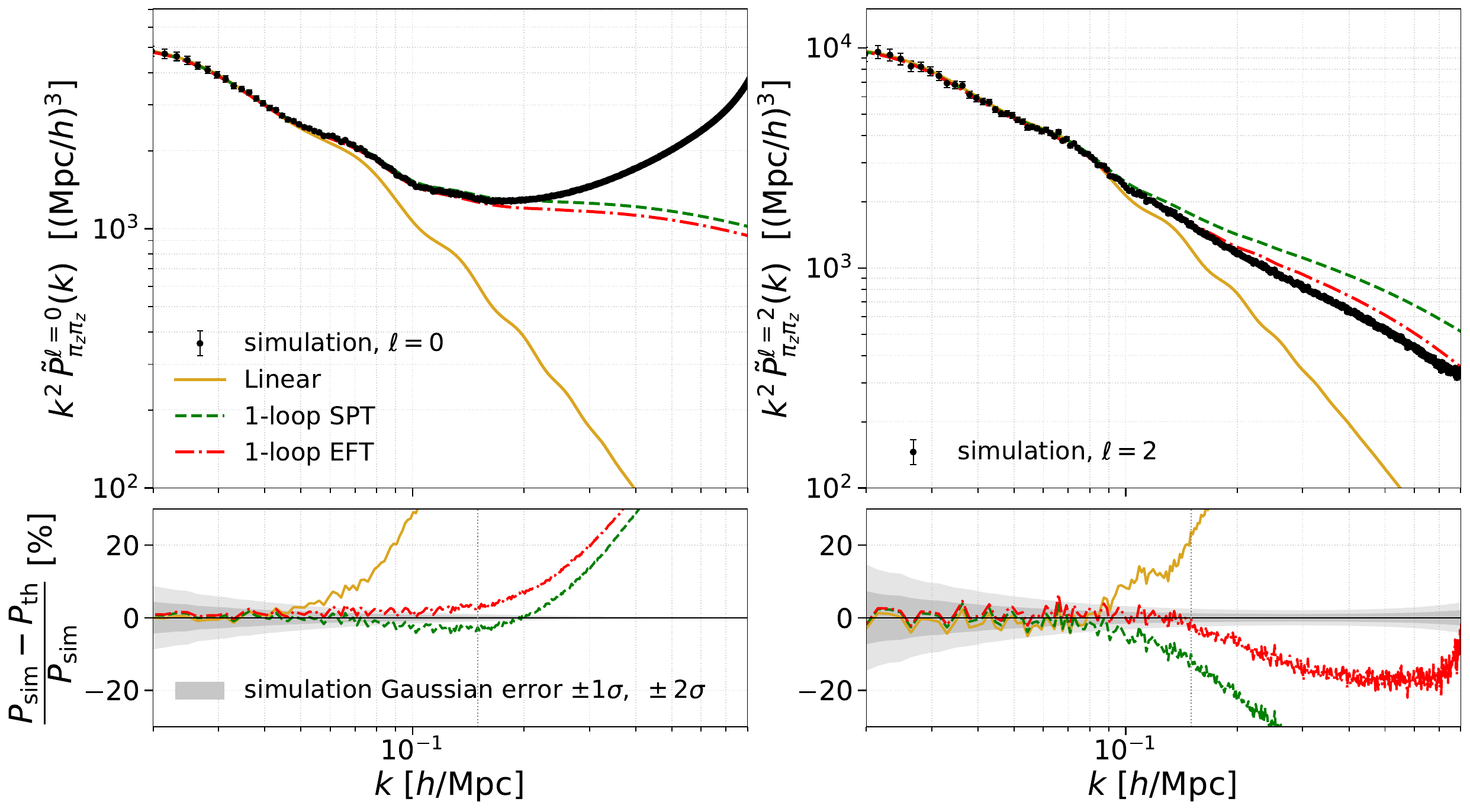}
    \caption{Comparison of \textit{Left:} $P_{\pi_z\pi_z}(k,\mu)$ monopole ($\ell=0$) and \textit{Right:} quadrupole ($\ell=2$) theoretical expressions against simulation. The expected improvement in matching simulation from tree-level (yellow) by adding the 1-loop correction (green, 1-loop SPT) and fitted counterterm (red, 1-loop EFT) is as expected.
    We further see that, as the momentum field is more non-linear, the range of validity of the theoretical model is further reduced. The counterterm coefficient was fitted to $P_{\pi_z\pi_z}^{\ell=2}$ up to $k=0.15\: \iM$. 
    }
    \label{fig:pk_pipi}
\end{figure}

We begin with the matter-matter and matter-momentum power spectra in Fig.~\ref{fig:pk_mm_mpi} followed by the momentum-momentum power spectrum in Fig.~\ref{fig:pk_pipi}.
Analogously to the bispectrum, we compute multipoles for each of these power spectra over the line of sight angle $\mu\equiv\hat{\k}\cdot\hat{\z}$,
\be 
P^{\ell}_{XY}(k)=\frac{2\ell+1}{2}\int_{-1}^1 d\mu~P_{XY}(k,\mu) \mathcal{L}_\ell(\mu)~\,,\quad X,Y=\{\delta,\pi_z\}\,.
\ee 

Given the specific dependence of each power spectrum expression in Section \ref{sec:pk} on $\mu$, we compute the monopole of the matter-matter power spectrum $P_{\delta \delta}^{\ell=0}$, the dipole of the matter-momentum power spectrum $P_{\delta\pi_z}^{\ell=1}$, and both the monopole and quadrupole of the momentum-momentum power spectrum $P_{\pi_z\pi_z}^{\ell=0, 2}$. Each measurement from simulation has an error-bar computed assuming a disconnected Gaussian variance (following~\cite{Chudaykin:2019ock}) and evaluated on the measured power spectrum multipoles.

Fig.~\ref{fig:pk_mm_mpi} shows the comparison of $P_{\delta\delta}$ and $P_{\delta \pi}$ between simulation and theory, demonstrating the expected structure of improvement between linear, 1-loop SPT, and 1-loop EFT theory. We also see that the matter-momentum power spectra presents a worse match against the simulation, expected as the momentum field is known to be more non-linear \cite{Baldauf:2015aha}. Figure \ref{fig:pk_pipi} shows the monopole and quadrupole moments of the $P_{\pi_z\pi_z}(k, \mu)$ power spectrum. Consistent with the previous claim about greater non-linearity in the momentum field, each theoretical $P_{\pi_z\pi_z}$ expression provides a narrower range of validity as compared to $P_{\delta \delta}$ and $P_{\delta\pi_z}$. For the momentum-momentum monopole, we also see that the linear theory expression fails quite drastically at a relatively low $k$. This can be explained by the presence of the transverse $\propto \mu^0$ correction 
in our Eq.~\eqref{eq:Pi1l},
which is also evident in the leading IR and UV limits of $P_{\pi_z\pi_z}$, Eq.~\eqref{eq:IRUVpipi}.\footnote{Previously it appeared in Equation 3.25 of \cite{vlah_2012}.}

\begin{figure}[H]
    \centering
    \includegraphics[width=0.9\linewidth]{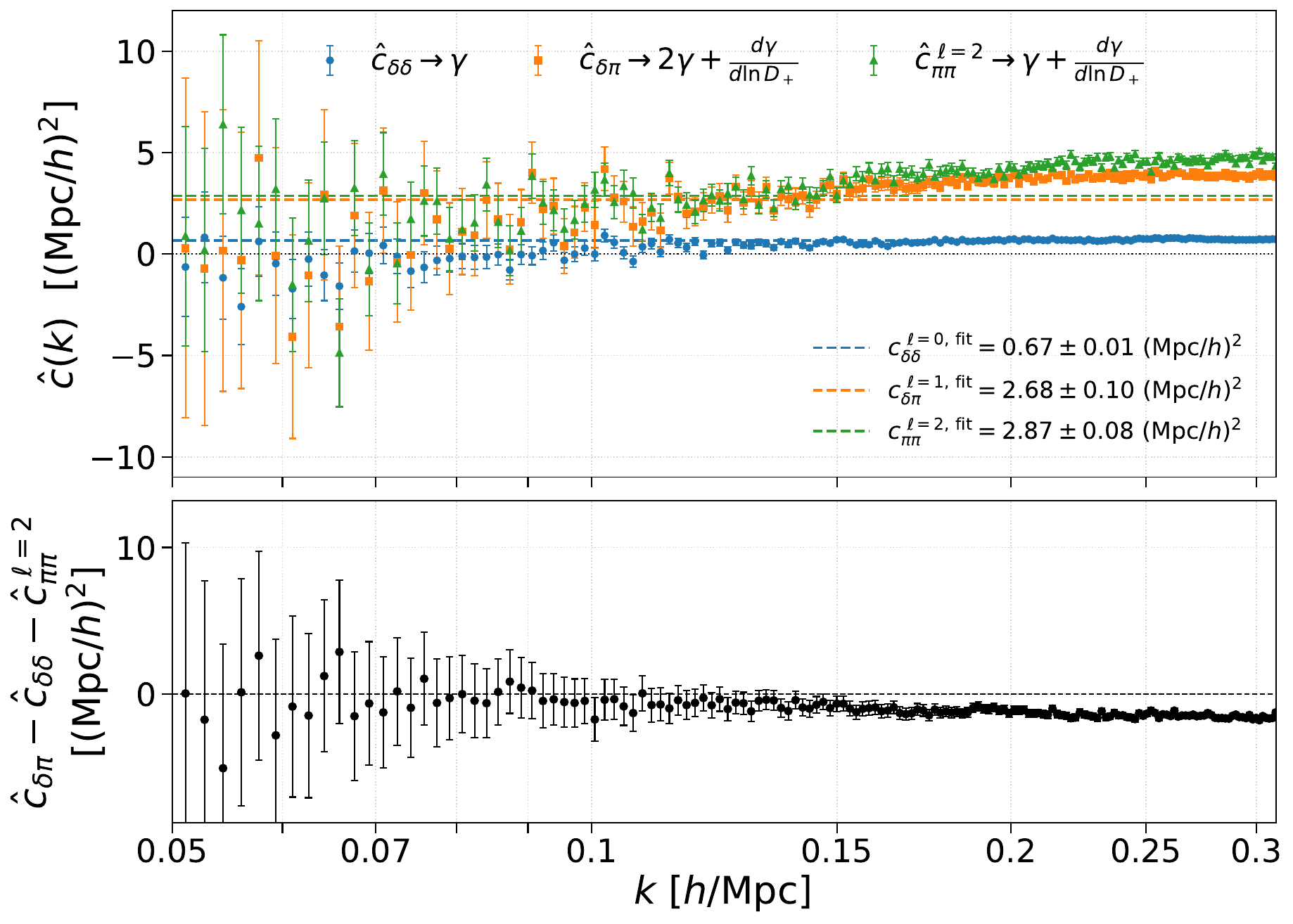}
    \caption{Counterterm coefficient estimators for $P_{\delta \delta}^{\ell=0}$, $P_{\delta \pi_z}^{\ell=1}$, $P_{\pi_z \pi_z}^{\ell=2}$, computed via the ratio of $P_{\text{sim}}^{\ell}(k)-P^{\ell}_{SPT}(k)$ to the counterterm shapes in Equation \eqref{eq:power_spec_ctrs}. The estimator acquiring a scale-dependence signals the presence of higher-loop corrections.
    In the top panel, one can see a relatively steady behavior for the $P_{\delta \delta}$, but both $P_{\delta \pi_z}$ and $P_{\pi_z \pi_z}$ quickly acquire scale dependence. This behavior is well known and documented by \cite{Baldauf:2015aha}. The bottom panel shows a combination of counterterm estimators that is
    exactly zero at one loop, such that deviation from zero indicates the presence of higher order loop corrections. Fitted values for counterterm coefficients used in Figures \ref{fig:pk_mm_mpi} and \ref{fig:pk_pipi} are also shown, computed via a least-squares fit up to $k_{\rm max}=0.25\: \iM$ for $P_{\delta\delta}$ and $k_{\rm max}=0.15\: \iM$ for both $P_{\delta\pi_z}$ and $P_{\pi_z\pi_z}$. Respectively, we obtain $0.67\pm0.01\:(\M)^2, 2.68\pm0.10 \:(\M)^2$, and $2.87 \pm0.08\: (\M)^2$.
    }
    \label{fig:power_spec_ctrs}
\end{figure}

Fig.~\ref{fig:power_spec_ctrs}
demonstrates the fits to counterterm coefficients for the various power spectra above. Shown are the estimators of the counterterm coefficients for $P_{\delta \delta}^{\ell=0}$, $P_{\delta \pi_z}^{\ell=1}$, and $P_{\pi_z \pi_z}^{\ell=2}$ as a function of $k$, given by the residual of 1-loop SPT and simulation power spectra normalized by the counterterm shape. The counterterm coefficients are computed via a least-squares fit, weighted by the error bars as described above. While the counterterm of $P_{\delta\delta}^{\ell=0}$ is stable with scale up to $k=0.3\:\iM$, those of $P_{\delta \pi_z}$ and $P_{\pi_z \pi_z}$ quickly acquire scale-dependence, indicating the presence of higher-loop corrections. The work by Ref.~\cite{Baldauf:2015aha} shows these correlation functions with the 2-loop correction, and this result is consistent with their findings. Namely, power spectra involving the momentum field have 2-loop corrections which appear at a scale similar to that of the leading order counterterm contribution. As such, we use relatively lower ranges of counterterm coefficient fitting for these two power spectra. Additionally, we have chosen only to use the quadrupole of $P_{\pi_z \pi_z}$ to fit the counterterm coefficient. Although this correction also shows up in the monopole, additionally present is the non-canceling 1-loop IR limit. This must be added in by hand as discussed in Section \ref{sec:uv}, which requires a choice of a UV cutoff. Practically, this would be fit as another free parameter, but for simplicity we neglect $P_{\pi_z\pi_z}^{\ell=0}$ in fitting the counterterm coefficient. 

\subsection{Comparison of $B_{mm\pi}^{\ell=1}(k_1, k_2, k_3)$ against Simulation}
Here, we present the comparison of the 1-loop theoretical bispectrum with the 3-d momentum field against the same bispectrum computed from the FLAMINGO L5p6 DMO simulation.
For the bispectrum covariance used in the plots and fitting counterterm coefficients, we use the disconnected Gaussian covariance, presenting the explicit expression in Appendix \ref{app:sigma_B}. 
Following~\cite{Ivanov:2021kcd,Ivanov:2023qzb}, 
we compute the covariance
in the Gaussian linear theory 
approximation. 

\begin{figure}[H]
    \centering
    \makebox[\linewidth][c]{\includegraphics[width=1.\linewidth]{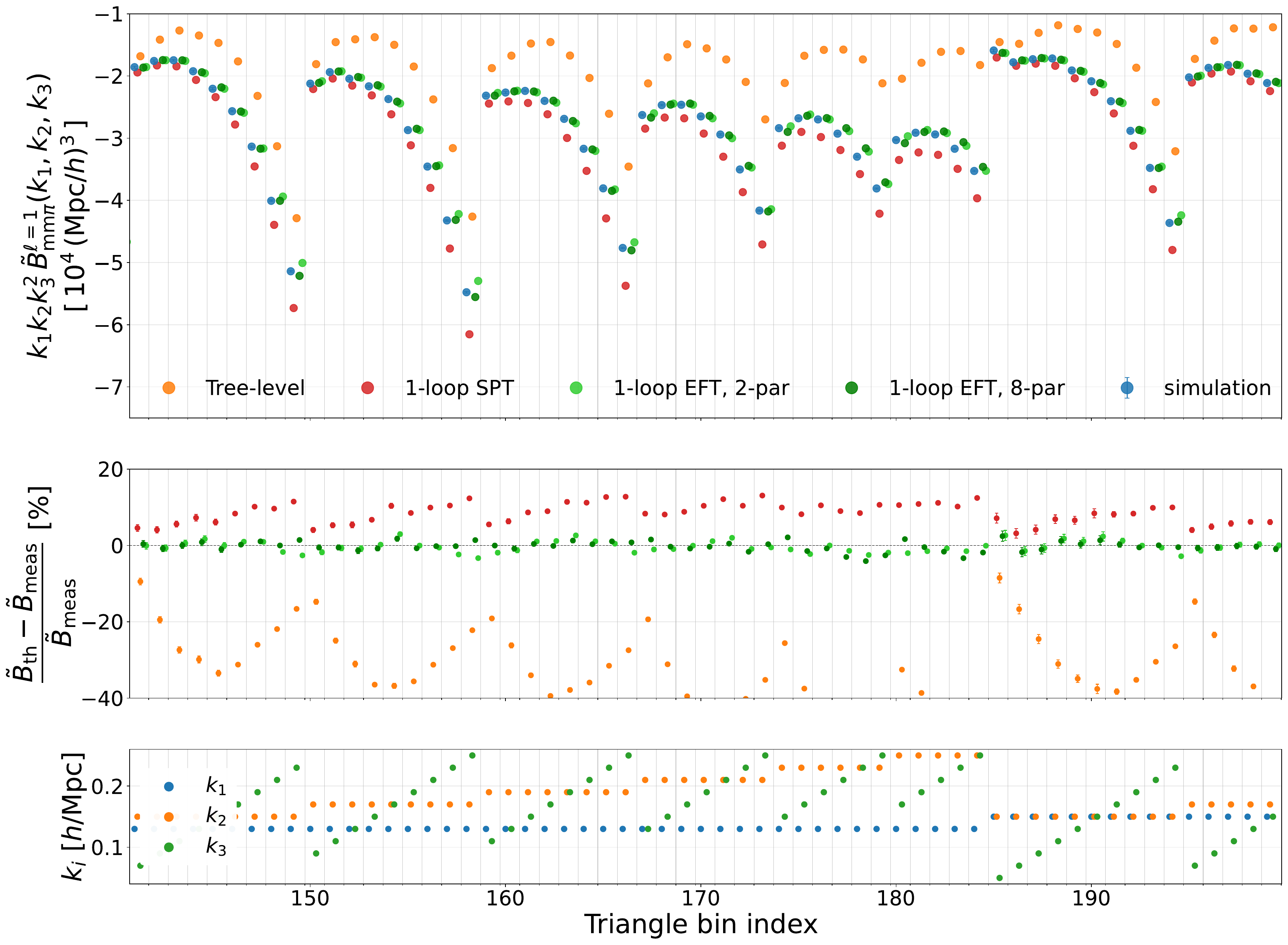}}
    \caption{$B^{mm\pi}_{\ell=1}$ comparison between theory and simulation for the raw estimator triangle ordering shown in the bottom panel. While lacking a coherent picture across different triangles, the improvement in matching the simulation from the tree-level (orange points) by adding the 1-loop correction (red points) and the counterterm corrections (light and dark green points) is evident. The fractional residual in the middle panel shows that tree-level theory consistently underestimates the simulation, which is resolved at the $\sim 10\%$ level just by the 1-loop correction. Including the two-parameter counterterm contribution (Equation \eqref{eq:ctr_time_dep_m}) with assumed time dependence to express counterterm coefficients in terms of $\gamma, m$, provides $<5\%$ accuracy to the simulation across triangles shown.
    }
    \label{fig:bispec_over_idxs}
\end{figure}

We begin by visualizing the different theoretical templates against the simulation bispectrum estimates. In practice, this is not so straightforward: there are thousands of triangle bispectrum evaluations with various $k_1, k_2, k_3$ and no clear way to order them in increasing non-linearity. To give a sense for this, we present the comparison between theory and simulation in the raw ordering of triangles used by the estimator in Figure \ref{fig:bispec_over_idxs}. Here we present a tiny subset of all the triangle configurations: 60 out of 3822. The bottom row shows these different triangle bins at which the theory and simulation are compared. These bins are ordered primarily in the increasing value of $k_1$, then $k_2$, and finally $k_3$ ranges between all allowed values that close the triangle/satisfy triangle conditions. We have chosen to only visualize triangles where all legs satisfy $k_i < 0.25 \: \iM$. 

\begin{figure}[H]
    \centering
    \makebox[\linewidth][c]{\includegraphics[width=1.\linewidth]{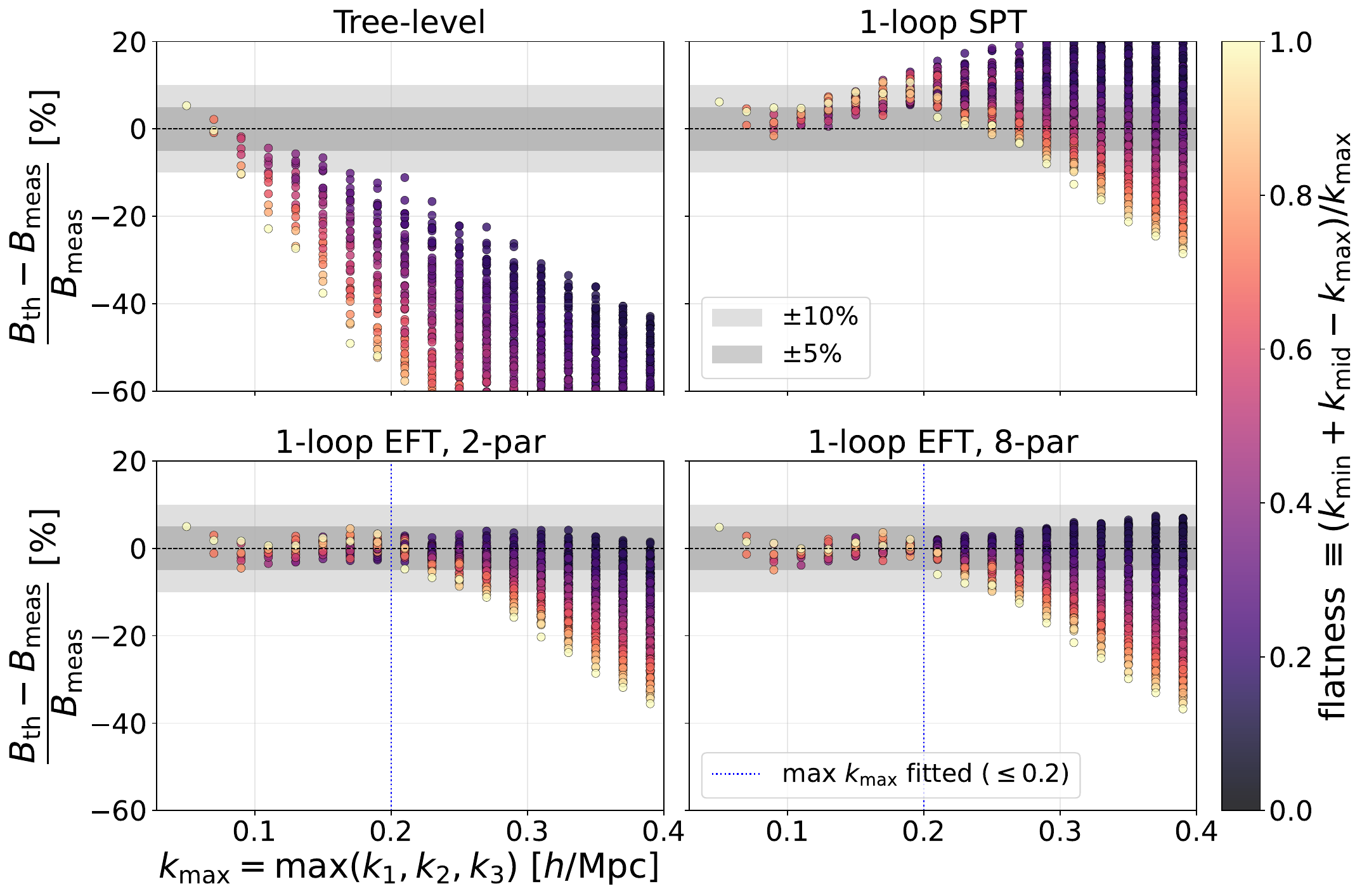}}
    \caption{Comparison of fractional deviation from simulation bispectrum measurements of various theoretical models. Shown are the tree-level (top left), 1-loop SPT (top right), and 1-loop EFT (bottom row) for two methods to fit the counterterm coefficients. At the level of $\sim 5 \%$ agreement we can roughly extract the maximum range to which these theoretical models match the simulation. Expectedly, tree-level is the worst, agreeing only at $k_{\rm max}= 0.07\:\iM$. The pure 1-loop correction pushes this range out to $k_{\rm max}= 0.11\:\iM$, and with either versions of the counterterm correction we have $k_{\rm max}= 0.23\:\iM$. Points are also colored by the flatness to give a rough sense of how non-linearity scales with triangle geometry. Squeezed configurations with flatness $\sim 0$ appear to be the most linear, with flattened triangles the most non-linear. The caveats to interpretations from this coloring are described above, namely that ordering $k_\text{max}$ is only sensitive to the longest triangle leg, ignoring the role of the shorter two legs. 
    }
    \label{fig:bispec_4_panel}
\end{figure}

We continue with the choice of $k_{\rm max}=\text{max}(k_1, k_2, k_3)$ to order triangle bins in terms of increasing non-linearity in Figure \ref{fig:bispec_4_panel}. This scheme is imperfect, as it neglects both the length of the shorter triangle legs, as well as the dependence of non-linearity on the shape of the triangle.
Nonetheless, it gives us a rough proxy, and aligns with the idea that perturbation theory allows us to model $\delta(\k)$ up to some $k_{\rm max}$, beyond which the perturbative assumptions break down. This choice of cutting on $k_{\rm max}$ also aligns with existing EFT bispectrum analyses \cite{Ivanov:2021kcd, bakx2025oneloopgalaxybispectrumconsistent}. 

\begin{figure}[H]
    \centering
    \makebox[\linewidth][c]{%
    \begin{minipage}[t]{0.5\linewidth}
        \centering
        \includegraphics[width=\linewidth]{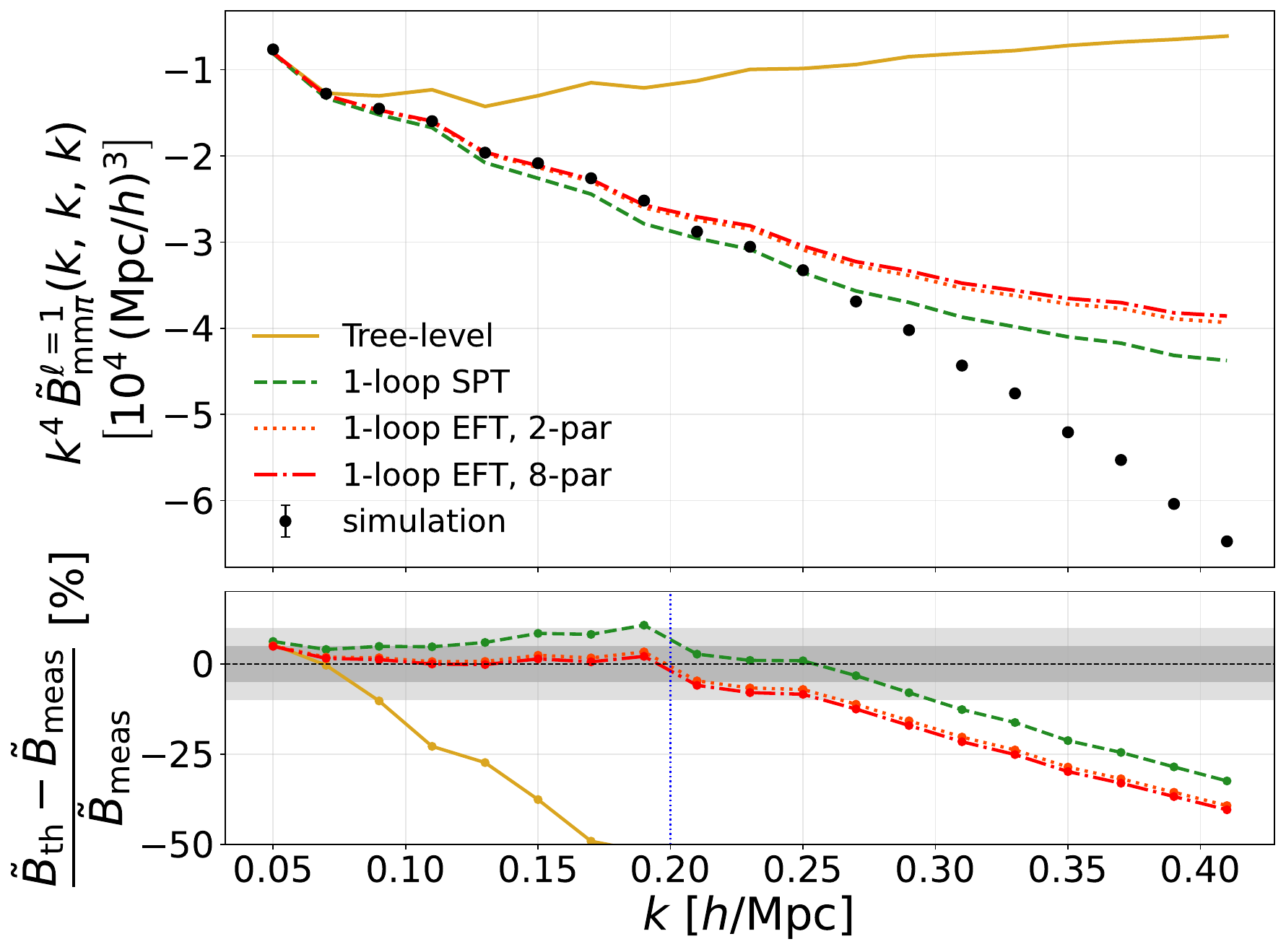}
        \caption{Comparison of simulation bispectrum measurement and theoretical models for equilateral triangles, i.e. $k_1 = k_2 = k_3=k$. At $5\%$ deviation from the simulation, we have roughly $k_{\rm max}=0.23\:\iM$ for 1-loop EFT. The lack of equilateral configurations past the Nyquist limit is also evident at $k=0.41\:\iM$. }
        \label{fig:bispec_equil}
    \end{minipage}\hspace{0.02\linewidth}%
    \begin{minipage}[t]{0.5\linewidth}
        \centering
        \includegraphics[width=\linewidth]{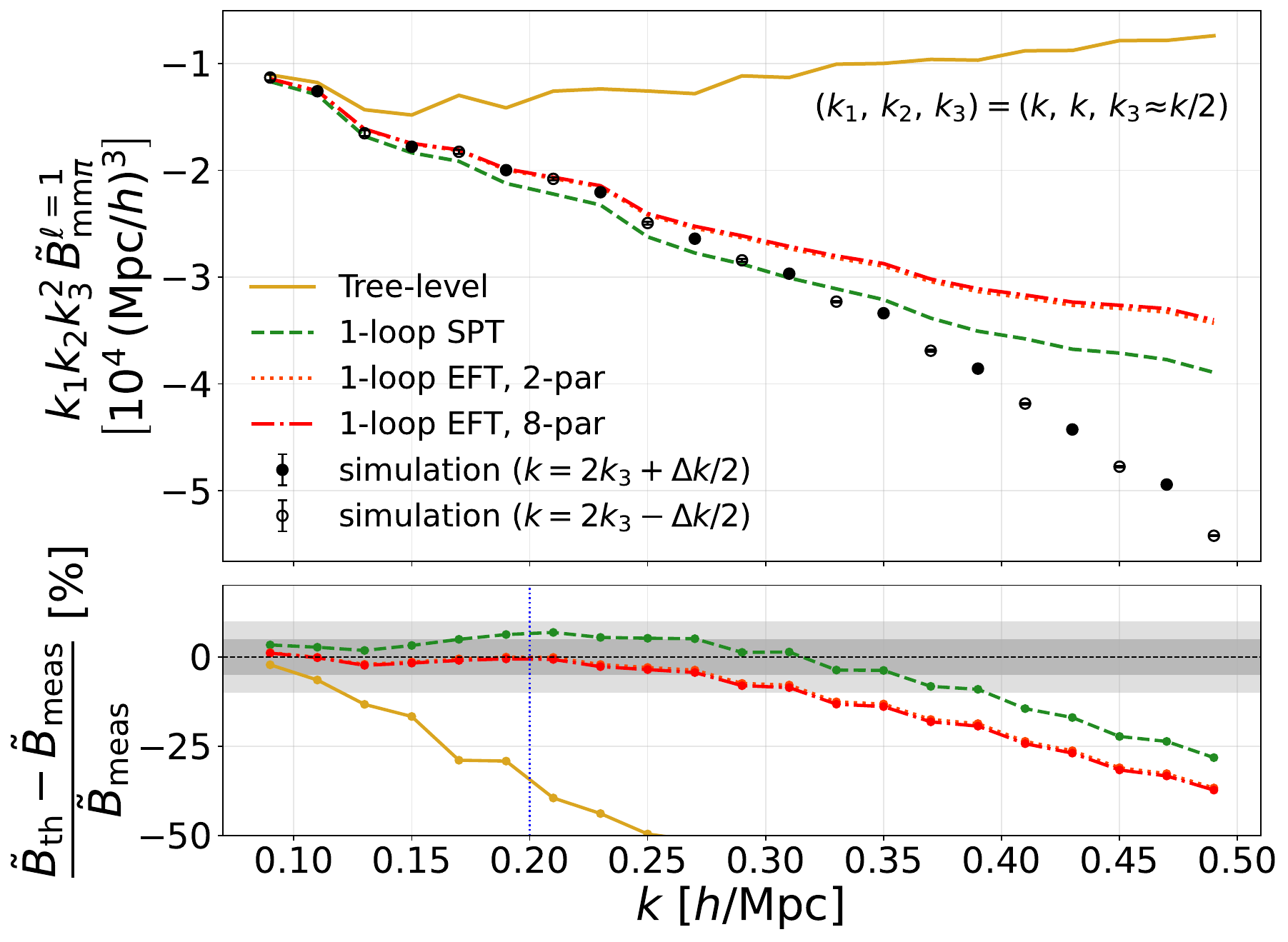}
        \caption{Comparison of simulation bispectrum measurement and theoretical models for isosceles triangles, with density field legs $k_1=k_2=k$ and momentum field leg $k_3\approx k/2$. At $5\%$ deviation from the simulation, we have roughly $k_{\rm max}=0.27\:\iM$ for 1-loop EFT. Since the choice of $(k_1, k_2, k_3)=(k, k, k/2)$ does not fall exactly on the bin centers, we show two sets of simulation evaluations with shaded $(k=2k_3+0.01\:\iM)$ and open $(k=2k_3-0.01\:\iM)$ circles. }
        \label{fig:bispec_isoc}
    \end{minipage}}%
\end{figure}

\begin{figure}[H]
    \centering
    \makebox[\linewidth][c]{%
    \includegraphics[width=0.5\linewidth]{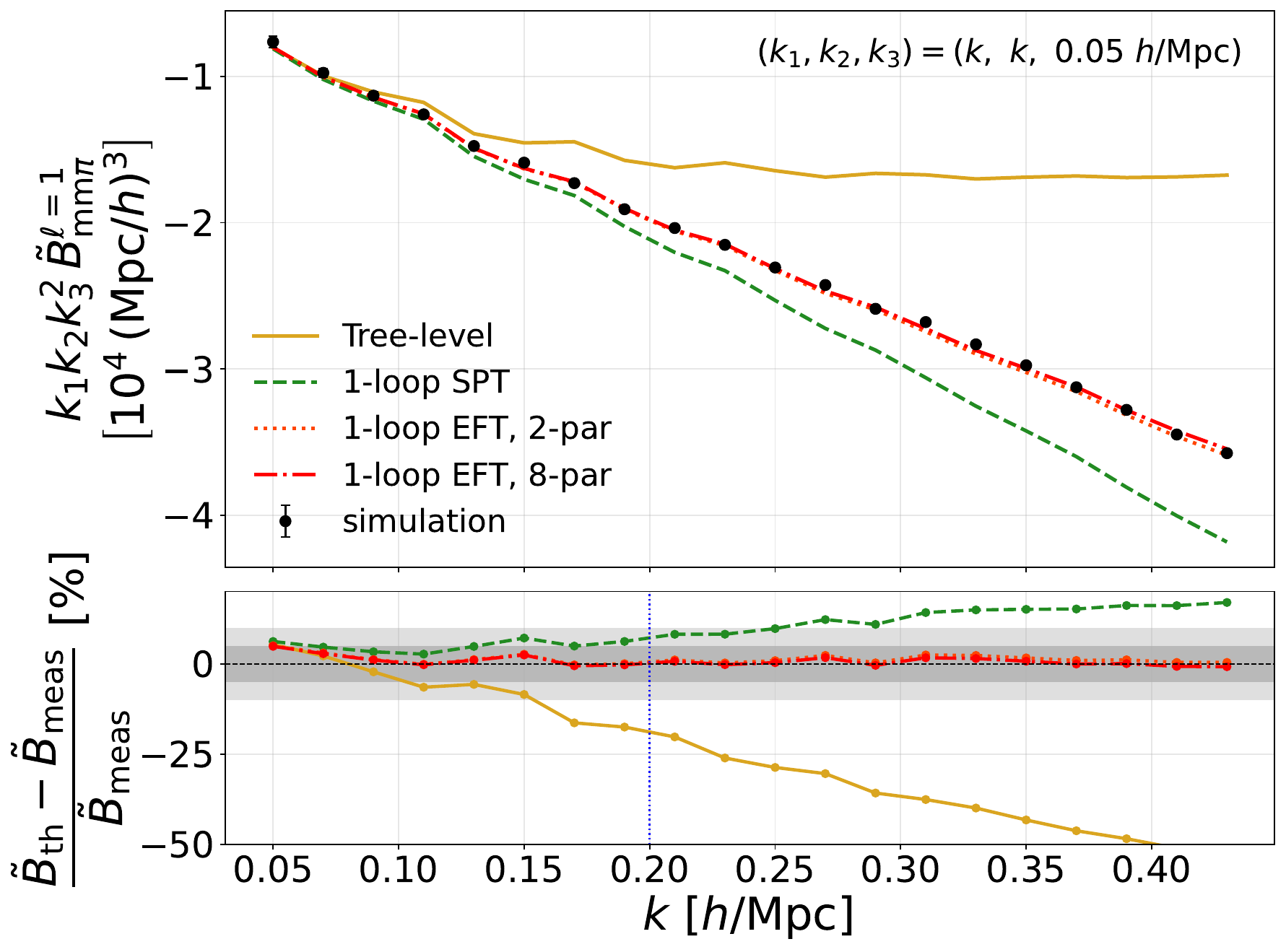}\hspace{0.02\linewidth}%
    \includegraphics[width=0.5\linewidth]{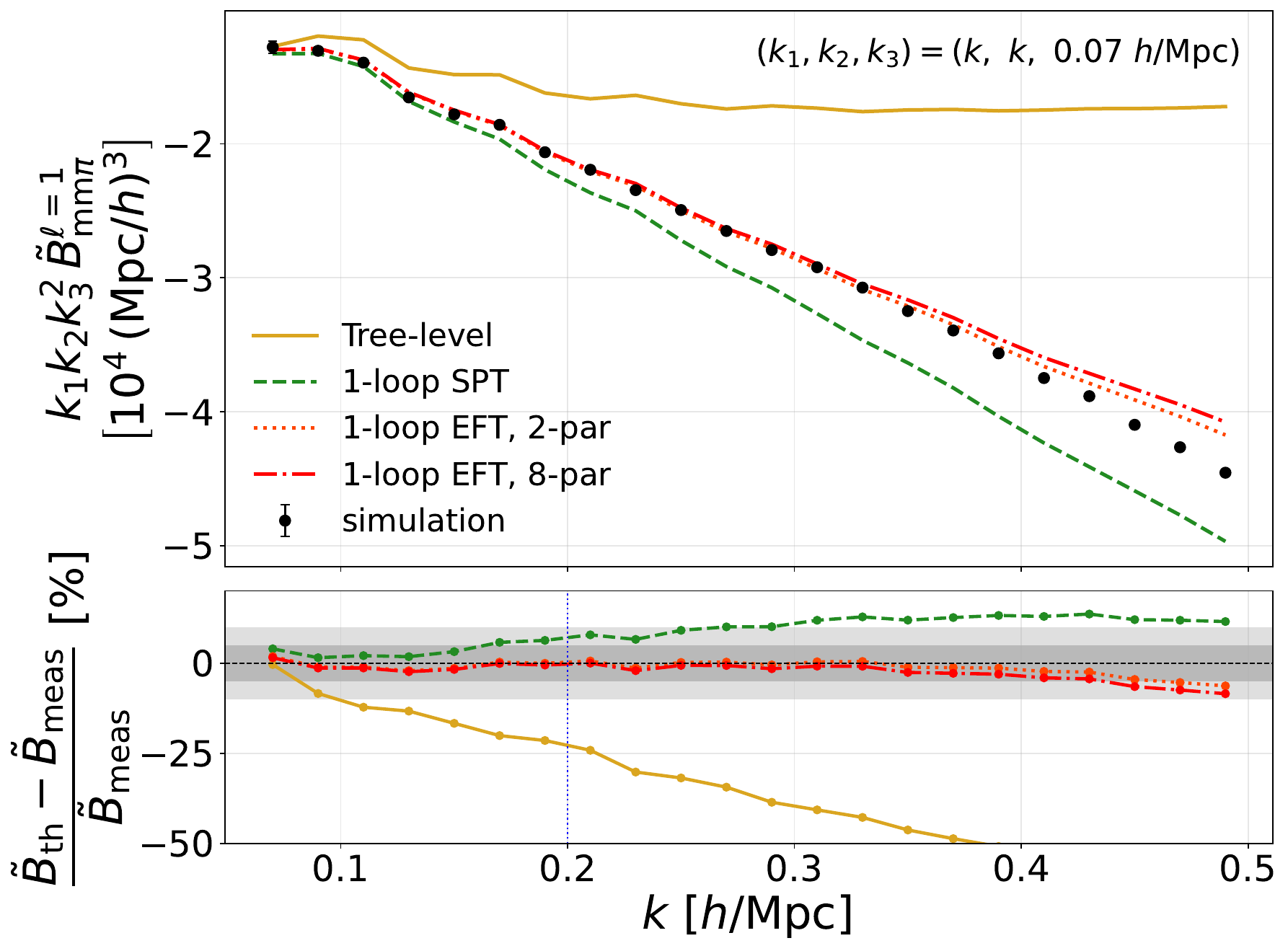}}
    \caption{Comparison of simulation bispectrum measurement and theoretical models for squeezed triangle configurations. \textit{Left:} the momentum field takes the short leg $k_3 = 0.05\:\iM$, \textit{Right:} $k_3 = 0.07\:\iM$. At $k_3=0.05 \:\iM$, 1-loop EFT agrees to $<5\%$ for the entire range of $k$ shown. At $k_3=0.07 \:\iM$, 1-loop EFT deviates past $5\%$ relative to the simulation at $k=0.43 \:\iM$. 
    }
    \label{fig:bispec_squeezed}
\end{figure}

For a cleaner visualization of the match between the simulation and theoretical bispectrum predictions, we consider three cases of fixed triangle geometry $(k_1, k_2, k_3)$: equilateral $(k, k, k)$ in Fig. \ref{fig:bispec_equil}, isosceles $(k, k, k/2)$ in Fig. \ref{fig:bispec_isoc}, and squeezed $(k, k, 0.05 \: \iM)$, $(k, k, 0.07 \: \iM)$ in Fig. \ref{fig:bispec_squeezed}. While these represent only a subset of all the triangle bins evaluated, most of which are scalene, each comparison presents a straightforward ordering in increasing non-linearity. These figures also serve to support the statement that the non-linearity of a given triangle configuration is strongly dependent on its geometry and cannot be easily captured by some function of the side lengths alone, such as $\text{max}(k_1, k_2, k_3)$. Taking the threshold of maximum range at the point where the disagreement of the full EFT theoretical model and simulation measurement exceeds $5\%$, the equilateral configuration yields $k_{\rm max}=0.23\:\iM$, the isosceles slightly better with $k_\text{max}=0.27\:\iM$, and the squeezed configurations much better: staying within $5 \%$ for the entire range of $k$ with the short leg $k_3=0.05\:\iM$.

\begin{figure}[H]
    \centering
    \makebox[\linewidth][c]{\includegraphics[width=1.\linewidth]{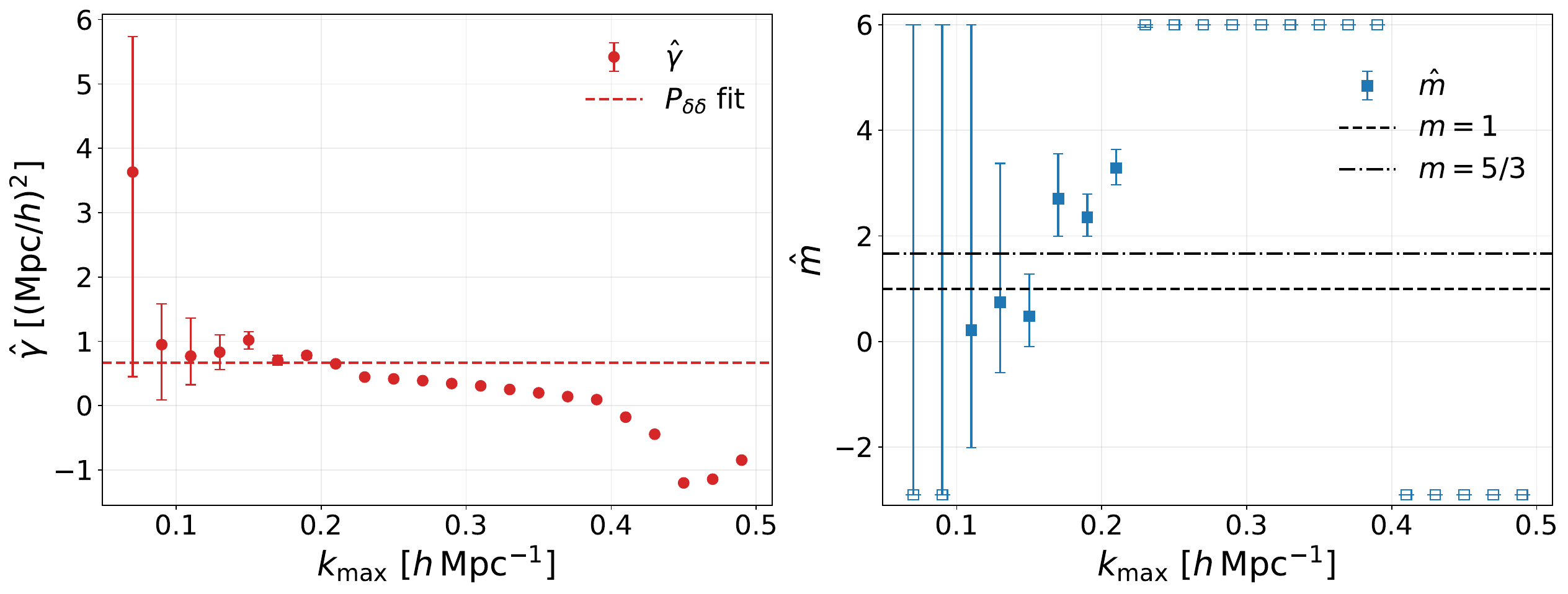}}
    \caption{Estimators for counterterm correction to the $B_{\delta\delta\pi_z}^{\ell=1}(k_1, k_2, k_3)$ bispectrum under the 2-parameter time dependence model. Each point corresponds to the best-fit value using only triangles in a single $k_{\rm max}$ bin, as opposed to the cumulative fit. The stability of the estimators holds to around $k_{\rm max}=0.2\:\iM$, motivating this as the fiducial choice for counterterm fitting. We find good agreement in the fitted value of $\gamma$ between power spectrum and bispectrum estimators, supporting the counterterm derivation. We also find that the power law scaling parameter $m$ is not very well constrained, but roughly agrees with proposed values, explicated in Appendix \ref{app:M_n_ctr}. Empty boxes in the estimator of $m$ indicate the fitted value falling outside the prior range, corresponding to a break-down of this simple time-dependence model. We also include horizontal lines of $m=1$ (UV-scaling) and $m=5/3$ (Lifshitz symmetry). Upon a least-squares fit up to $k_{\rm max}=0.2 \: \iM$, we obtain the values $\gamma=0.78 \pm 0.05\:(\M)^2$ and $m=2.35\pm 0.4$.}
    \label{fig:bispec_2par_m_gamma}
\end{figure}

To fit the counterterm graph coefficients we follow a similar mechanism to that for the power spectrum. First, a $k_{\rm max}$ is chosen, and the fit is performed for all triangle configurations satisfying $\text{max}(k_1, k_2, k_3)\leq k_{\rm max}$. We choose $k_{\rm max}=0.25 \: \iM$ as the regime where further increasing its value leads to overfitting for high $k$ at the expense of low $k$ configurations. With the counterterm correction shapes given in Equations \ref{eq:F_2_ctr}, \ref{eq:M_2_ctr}, the counterterm coefficients described in Appendix \ref{app:M_n_ctr} are optimized to minimize the quadratic error between the simulation and 1-loop corrections using the Gaussian covariance as an error estimate. We compare two different schemes for these counterterm coefficients. In the first, denoted 2-parameter, we assume the time-scaling of counterterm coefficients as a power law in the linear growth factor, e.g. $c_s^2(\tau) \propto D^{m+1}(\tau)$. As described in Appendix~\ref{app:M_n_ctr}, by using the appropriate Green's functions, the time scaling of each counterterm coefficients can be determined. By matching the counterterm shape to cancel the leading UV limit of the appropriate 1-loop correction, we establish consistency relations between these counterterm coefficients in terms of the power law parameter $m$. Additionally, $c_s^2$ or, equivalently, $\gamma$ appears as the second free parameter, scaling all the loop corrections equally. In the second counterterm scheme, we do not assume any kind of time-dependence relations, and fit all counterterm coefficients individually. Similar to Figure \ref{fig:power_spec_ctrs} for the power spectra, we present the estimated counterterm coefficients as a function of triangle $k_{\rm max}$ bin for the 2-parameter model with $\gamma, m$ in Figure \ref{fig:bispec_2par_m_gamma}. We find reasonable stability of fitted coefficients around $k_{\rm max}=0.2\:\iM$, supporting the statements of agreement with simulation. 

\subsection{$B_{\text{mm}\pi}^{\ell=1}$ for the kSZ Field}
The majority of this paper deals with the generalized case of the $\langle \delta \delta \pi_z \rangle$ bispectrum involving the full 3-dimensional momentum field, enabling a precise comparison of simulation measurement with low cosmic variance against theoretical predictions. While the majority of details for the application of the 1-loop bispectrum corrections to measuring the kSZ field are left to the follow-up paper, we present this short subsection as an introduction. We begin with the 3-dimensional momentum field bispectrum, presenting a comparison to the existing mechanism for measuring the kSZ signal by \cite{Smith26PhRvD:ksz_bisp}. We follow with the match of theoretical expression against simulation, now using the 2-dimensional projected momentum field and the corresponding transverse counterterm coefficient fits. 

\begin{figure}[H]
    \centering
    \makebox[\linewidth][c]{%
    \includegraphics[width=0.5\linewidth]{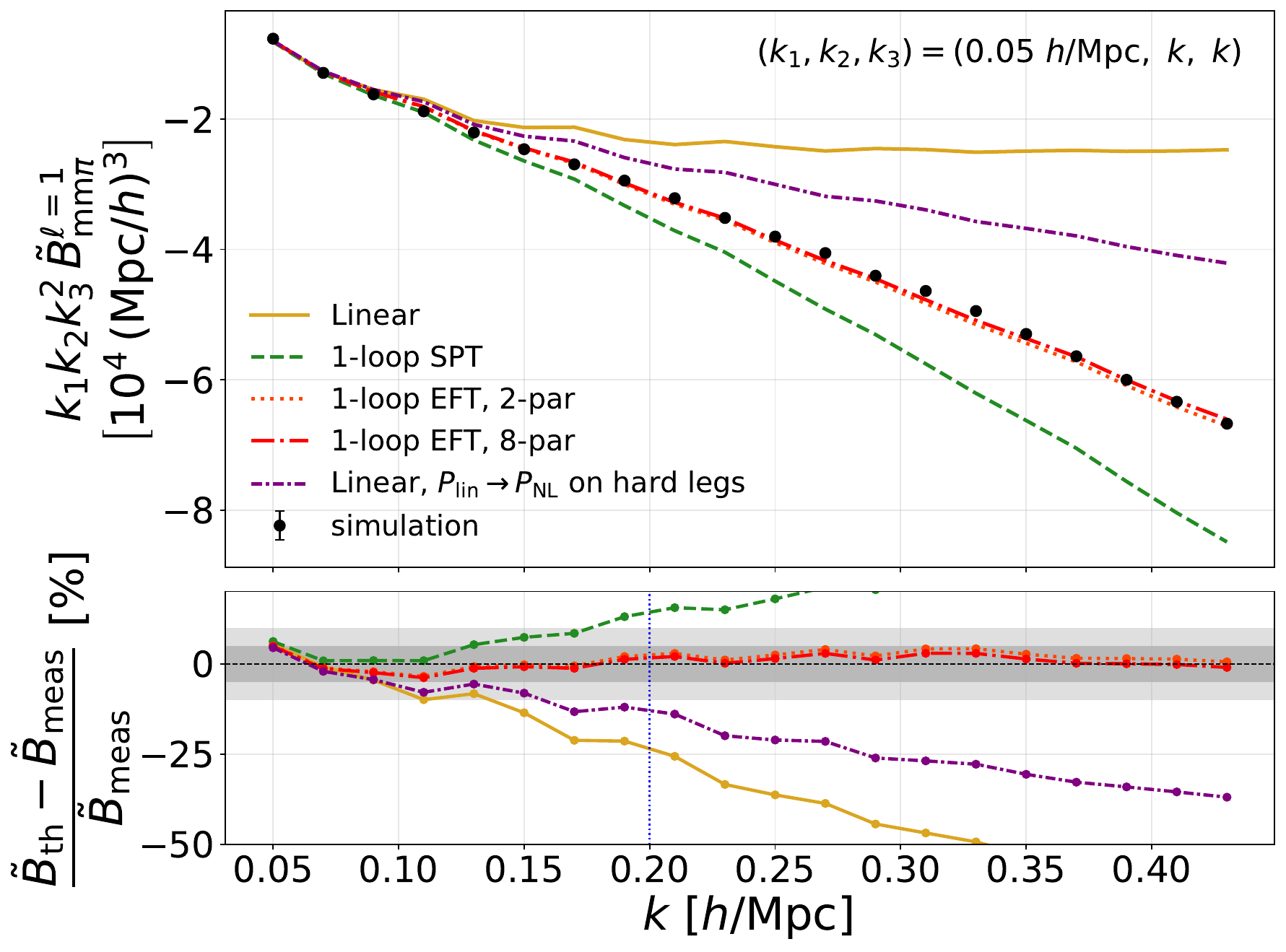}\hspace{0.02\linewidth}%
    \includegraphics[width=0.5\linewidth]{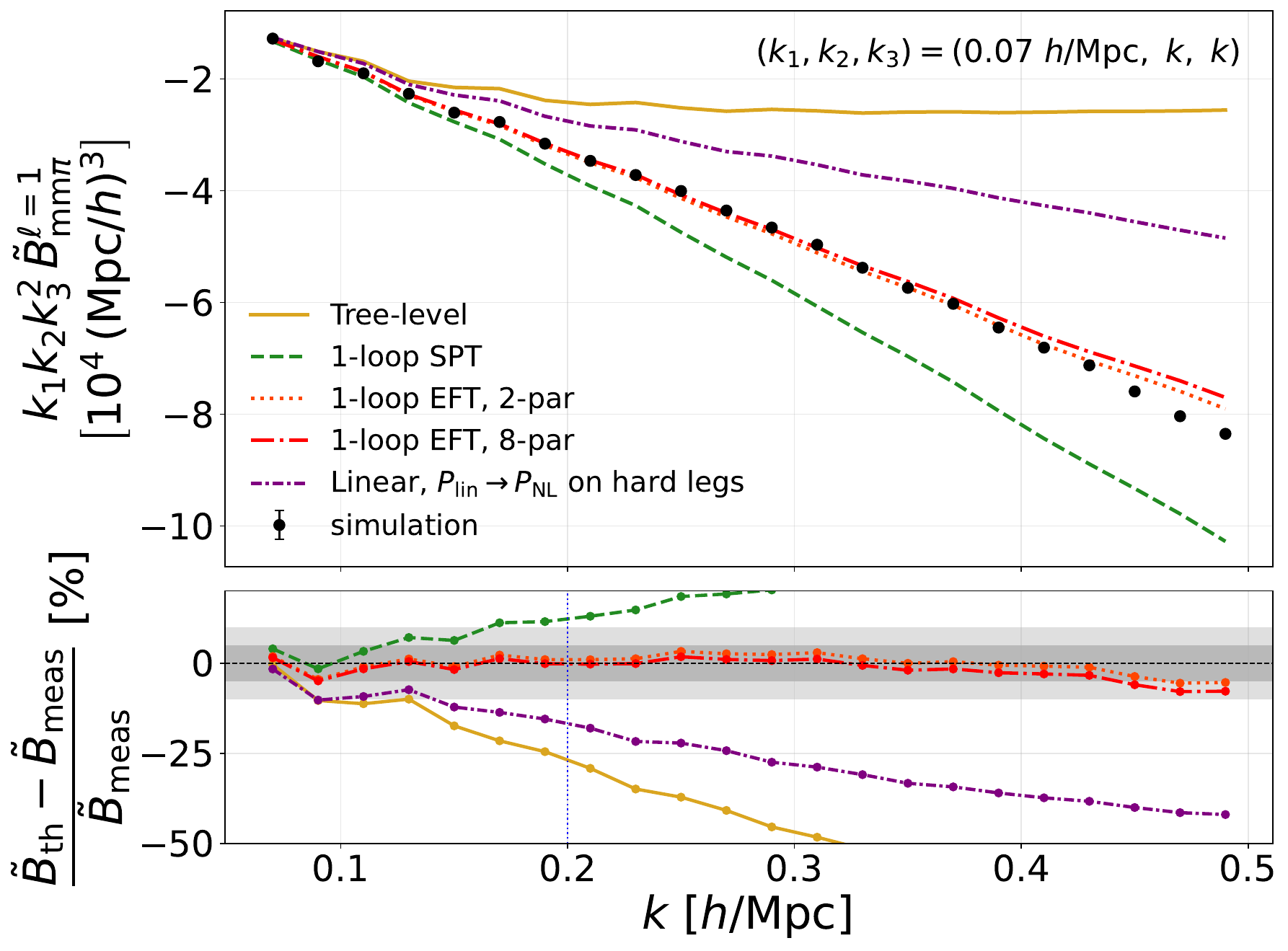}}
    \caption{Comparison of simulation vs. theoretical models for the squeezed triangle configuration, now with the momentum field $k_3$ as one of the long legs (\textit{left:} $k_3 = 0.05\:\iM$; \textit{right:} $k_3 = 0.07\:\iM$). Both 1-loop counterterm predictions agree with the simulation to $5\%$ for the entire k-range plotted. Additionally shown is the squeezed limit prediction of \cite{Smith26PhRvD:ksz_bisp} by replacing $P_{11}$ with $P_{\rm NL}$ in purple, demonstrating a slight improvement over the tree-level theory. }
    \label{fig:bispec_with_Pnl}
\end{figure}

In Figure \ref{fig:bispec_with_Pnl}, we make a similar comparison to Figure \ref{fig:bispec_squeezed} but with the momentum field as one of the non-linear high-$k$ legs and one of the density field taking the linear low-$k$ leg. We have also plotted the bispectrum prediction using the squeezed-limit assumption of \cite{Smith26PhRvD:ksz_bisp}, emphasizing that we are working with dark matter field, as opposed to electron and galaxy fields used by \cite{Smith26PhRvD:ksz_bisp}. Here, the tree-level prediction is modified by replacing the linear power spectrum $P_{11}(k_S)$ of the high-$k$ mode with the non-linear power spectrum $P_{\text{NL}}(k_S)$. We use the \texttt{camb} Halofit non-linear power spectrum for the same D3A cosmology at the snapshot redshift $z=0.5$ \cite{Takahashi:2012em}, which we have checked agrees with the simulation measured $P_{\text{mm}}$ over all scales considered. As described by \cite{Smith26PhRvD:ksz_bisp}, this approximation for the squeezed-limit bispectrum is equivalent to other formalisms used to measure and derive information from measurements of the kSZ field. While the short legs (long mode) of the triangles in Figure \ref{fig:bispec_squeezed} are $k=0.05\:\iM, 0.07\:\iM$ respectively, which could be chosen as longer physical modes, this demonstrates a potential source of bias in current analyses of the kSZ field.

\begin{figure}
    \centering
    \makebox[\linewidth][c]{\includegraphics[width=1.\linewidth]{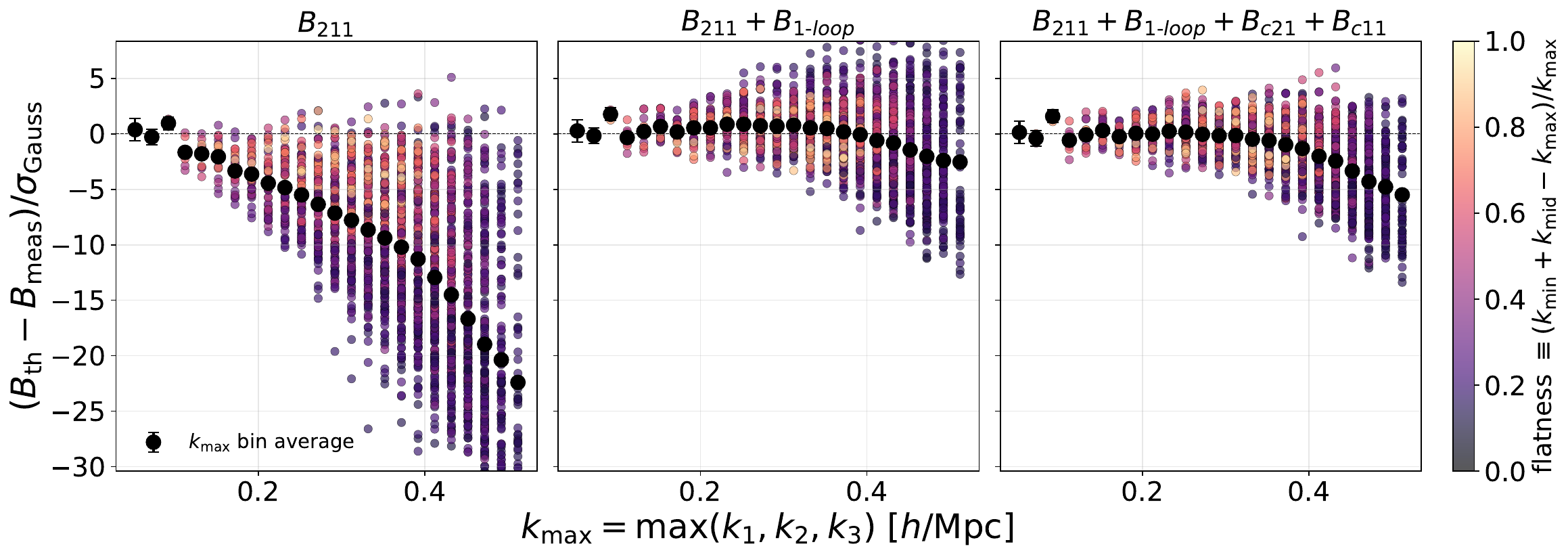}}
    \caption{Comparison of L11p2 simulation bispectrum measurements against various theoretical models --- \textit{left:} tree-level, \textit{center:} 1-loop SPT, \textit{right:} 1-loop EFT. The bispectrum dipole is now computed over $\mu_1$, as the projected momentum field has identically $\mu_3=0$. The deviation of tree-level theory is clear, as well as the large extension in fitting range of validity from adding 1-loop and counterterm corrections. Black points indicate the difference between simulation and theoretical prediction averaged over triangles under a fixed $k_{\rm max}$ to visually reduce the spread induced by sample variance. As discussed in Section~\ref{sec:uv}, the auto-correlation of the projected momentum field, necessary to compute the disconnected Gaussian variance, is zero at leading order. As such, we use the measured auto power-spectrum of the projected momentum field in computed error bars.}
    \label{fig:bispec_mu3=0}
\end{figure}

\begin{figure}[H]
    \centering
    \makebox[\linewidth][c]{\includegraphics[width=\linewidth]{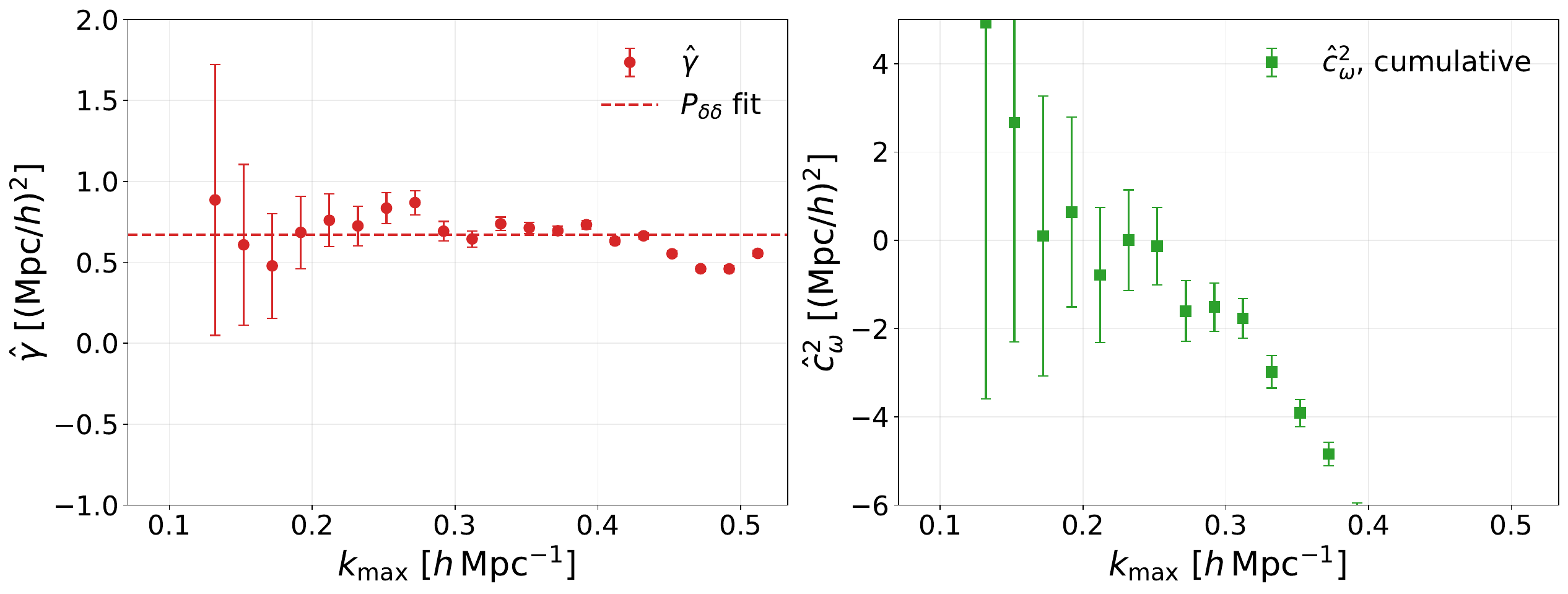}}
    \caption{Estimators for counterterm corrections to the kSZ-like bispectrum involving the projected momentum field, i.e. $\mu_3=0$, for the two surviving counterterm coefficients: $\gamma$ (left panel) and $\hat{c}^2_{\omega}$ (right panel). We find good agreement in $\gamma$ against the value found in the power spectrum and 3-d momentum field bispectrum. We show the fitted value of $c_\omega^2$ as a function of $k_{\rm max}$, as a cumulative fit for all triangles, i.e. $\leq k_{\rm max}$. Taking the extent to which coefficients aren't scale-dependent at $k_{\rm max}= 0.32\: \iM$, we demonstrate a first detection of uniquely the $c_\omega^2$ transverse counterterm correction. Using least-squares to fit these two coefficients up to $k_{\rm max}=0.32 \: \iM$, we obtain $\gamma=0.72 \pm0.03 \:(\M)^2$ and $\hat{c}^2_{\omega}=-1.77 \pm0.45 \:(\M)^2$.
    }
    \label{fig:bispec_projected_ctr_fit}
\end{figure}

To demonstrate the 1-loop EFT prediction for the case of the 2-d projected momentum field relevant to the case of kSZ measurement, we present Figure \ref{fig:bispec_mu3=0}.
As compared to Figure \ref{fig:bispec_4_panel}, there is significantly more scatter in the residual of theoretical predictions and simulation measurement sourced from the noisier bispectrum measurement. This can roughly be understood via the projected momentum field having $N_{\rm grid}^2$ modes as opposed to $N_{\rm grid}^3$. Averaging over the disagreement by $k_{\rm max}$ bin allows a somewhat consistent comparison between each theoretical model to be made. Similar to Figures \ref{fig:power_spec_ctrs} and \ref{fig:bispec_2par_m_gamma}, we visualize the fit of the two transverse counterterm coefficients appearing for the case of the projected momentum field in Figure \ref{fig:bispec_projected_ctr_fit}. Again, we find consistency with the power spectrum and 3-dimensional momentum field bispectrum in the value of $\gamma$, and make a $\approx 4\sigma$ detection of the transverse counterterm, $\hat{c}^2_{\omega}=-1.77 \pm0.45 \:(\M)^2$. 

\section{Discussions and Conclusions \label{sec:conc}}

In our work, we have demonstrated the accuracy of the EFT framework applied to modeling the $\langle \delta \delta \pi_z \rangle$ bispectrum, which can in turn be used for precision measurements of the kSZ effect. We began at the power spectrum level, quantifying the smallest scale $k_{\rm max}$ at which theoretical predictions agree with simulations to $5\%$. Across the power spectra considered, we found that the EFT models at 1-loop order increased $k_{\rm max}$ by a factor of roughly $2-3$. As the number of available modes is cubic with $k_{\rm max}$, this represents a large improvement in the amount of data which can be accurately modeled. 

We continued with the $\langle \delta \delta \pi_z \rangle$ bispectrum, deriving the four one-loop corrections, establishing a framework for efficient evaluation by turning integrals into sums over hypergeometric functions. To renormalize UV divergences, we computed the relevant counterterm expressions finding 9 free coefficients, introducing a novel transverse counterterm correction to the momentum field. We compared this theoretical prediction against binned bispectrum multipoles from a simulation via a Scoccimarro estimator, quantifying the smallest scale $k_{\rm max}$ for which all triangles with legs $k_i \leq k_{\rm max}$ agree with simulation to $5\%$. In Figure \ref{fig:bispec_4_panel} we found that the tree-level theory has $k_{\rm max}=0.07 \:\iM$, and the EFT 1-loop pushes out to $k_{\rm max}=0.23 \:\iM$. In the bispectrum, the number of triangles available scales $\propto k_{\rm max}^3$, with each triangle roughly containing $k_1 k_2 k_3 \sim k^3$ modes. Roughly, this implies that the amount of information in the bispectrum scales $\propto k_{\rm max}^6$; an increase of $k_{\rm max}$ from $0.07\:\iM$ to $0.23\:\iM$ with the EFT 1-loop model is therefore a huge gain. In fitting the counterterm contributions for the bispectrum dipole, we found that the transverse contribution is zeroed, leaving 8 free counterterms. Under both this fully-free 8 parameter model, as well as the simpler 2-parameter model assuming a time-dependence of the counterterm coefficients, we found a strong improvement in agreement with the simulation. In inspecting the fitted value of the 2-parameter model coefficients as a function of scale, we found consistency with the fitted value of $\gamma$ against $P_{\delta \delta}$, and further motivated the range of validity of the EFT 1-loop calculation. 

As non-linearity cannot be well captured by solely the longest triangle leg, we also looked at the agreement between theoretical predictions and simulation for three specific triangle configurations $(k_1, k_2, k_3)$: equilateral $\sim (k, k, k)$, isosceles $\sim (2k, 2k, k)$, and squeezed $\sim (k, k, 0.05 \:\iM), (k, k, 0.07 \:\iM)$. Among the three, we found that the equilateral case had the worst agreement at a given $k$, isosceles slightly better, with the squeezed configurations by far the best. While these fixed configurations contain a small subset of the available triangles to measure, a particular cosmological analysis, e.g. for primordial non-Gaussianity, may be primarily sensitive to different triangle geometries (see e.g.~\cite{Cabass:2022wjy,Cabass:2022ymb,Cabass:2022epm,Chudaykin:2025vdh} for the galaxy bispectrum case). In such a case, there is room for additional optimization of $k_{\rm max}$ for particular configurations that drive the signal to noise. 

While to make the most accurate comparison between theoretical expressions and simulations we used the full 3-dimensional momentum field, we presented the case of the 2-dimensional projected momentum field bispectrum relevant to kSZ observation in Figure \ref{fig:bispec_mu3=0}. 
While higher cosmic variance makes interpretation of this comparison less straightforward, by averaging over the disagreement between theory and simulation for different triangle configurations at a given $k_{\rm max}$ we see a similar picture to before: tree-level theory begins to fail before $k_{\rm max}=0.1 \:\iM$, and the EFT 1-loop model allows an extension to $\sim k_{\rm max}=0.32\:\iM$. Under projecting the momentum field, all longitudinal counterterms disappear from the bispectrum dipole, leaving only 2 parameters --- $\gamma$ and $c_\omega^2$. We found agreement in the fitted value of $\gamma$ to $P_{\delta\delta}$, and by using the larger L11p2 FLAMINGO box were able to present the first $\approx 4\sigma$ detection of the contribution of the transverse counterterm $c_\omega^2$. 

Since $c_\omega^2$ is generated by the transverse part of the velocity 
field, a non-zero $c_\omega^2$
also 
implies a detection of dark matter vorticity. It is well-known that vorticity
is absent to all orders in 
Standard Perturbation Theory~\cite{Bernardeau:2001qr}, 
but is ultimately present
in the dark matter simulations~\cite{Pueblas:2008uv}.
In the context of EFT, vorticity is generated by the 
dark matter stress-tensor
at next-to-leading order. 
Our detection of $c_\omega^2$ thus
confirms this non-trivial EFT prediction. 

Future papers in the current series will present EFT computations of the bispectrum $\langle \delta_g \delta_g \pi^e_z \rangle$ in the more realistic setting of a full galaxy bias expansion and will include the relevant associated complications (additional non-linear bias terms, large stochastic contributions from galaxy size and discreteness, and independent renormalization conditions for the density and momentum fields beyond the continuity equation) as well as the developments necessary for constraining cosmological parameters, such as the amplitude of local primordial non-Gaussianity ($f_\mathrm{NL}$).

\acknowledgments
JMS also acknowledges that support for this work was provided by The Brinson Foundation through a Brinson Prize Fellowship grant.
This research has made use of NASA's Astrophysics Data System.
Generative AI has been used to assist in coding and validating computations, and has not been used to write any text in this paper.
This work has made heavy use of the SubMIT cluster at MIT. 
This work is supported by NSF grant 2008031.
\appendix

\section{Explicit Momentum Field kernels $M_n$ \label{app:M_n}}
Here we provide the explicit forms for the unsymmetrized momentum-field kernels from Eq. \eqref{eq:M_n_sum}. Note that the expressions are not written with the symmetrization over arguments, which must be done prior to evaluating expressions.
\begin{align}
M_{1}(\boldsymbol{q}_{1}) &= f\HH \frac{i \mu_1}{q_1}, \notag\\[4pt]
M_2^{\text{non-symm.}}(\boldsymbol{q}_{1},\boldsymbol{q}_{2})
  &= f\HH \left[\,\frac{i \mu_{12}}{q_{12}}G_2(\q_1, \q_2)+ \frac{i\,\mu_{2}}{q_{2}} \right], \notag\\[6pt]
M_{3}^{\text{non-symm.}}(\boldsymbol{q}_{1},\boldsymbol{q}_{2},\boldsymbol{q}_{3})
  &= f\HH \left[\frac{i \mu_{123}}{q_{123}}G_3(\q_1, \q_2, \q_3)+
        \frac{i\,\mu_{1}}{q_{1}}\, F_{2}(\boldsymbol{q}_{2},\boldsymbol{q}_{3})
      + \frac{i\,\mu_{23}}{q_{23}}\, G_{2}(\boldsymbol{q}_{2},\boldsymbol{q}_{3})
     \right], \notag\\[6pt]
M_{4}^{\text{non-symm.}}(\boldsymbol{q}_{1},\boldsymbol{q}_{2},\boldsymbol{q}_{3},\boldsymbol{q}_{4})
  &= f\HH \bigg[ \frac{i\mu_{1234}}{q_{1234}}G_4(\q_1, \q_2, \q_3, \q_4)+
        \frac{i\,\mu_{1}}{q_{1}}\, F_{3}(\boldsymbol{q}_{2},\boldsymbol{q}_{3},\boldsymbol{q}_{4})\notag \\
        & \qquad \quad+ \frac{i\,\mu_{234}}{q_{234}}\, G_{3}(\boldsymbol{q}_{2},\boldsymbol{q}_{3},\boldsymbol{q}_{4})
      + \frac{i\,\mu_{34}}{q_{34}}\, G_{2}(\boldsymbol{q}_{3},\boldsymbol{q}_{4})\, F_{2}(\boldsymbol{q}_{1},\boldsymbol{q}_{2})
     \bigg].
\end{align}

\section{$F_{2}^{\text{ctr}}$ and $M_{2}^{\text{ctr}}$ \label{app:M_n_ctr}}
Here, we fill in details of the second-order counterterm kernels, $F_{2}^{\text{ctr}}$ and $M_{2}^{\text{ctr}}$. These are necessary to renormalize the 1-loop $\langle\delta \delta \pi_z\rangle$ bispectrum, as described in Section~\ref{sec:uv}. We begin with the 
EFT fluid equations:
\be 
\begin{split}
& \partial_\tau\delta+\partial_i\pi^i=0,
 \\
  &\partial_\tau v^i+\HH v^i+\partial^i\phi+v^j\partial_jv^i=\tau^i_\textit{vis}\,,
 \label{eq:exact}
\end{split}
\ee 
where $\pi^i\equiv(1+\delta)v^i$.
Using the transverse wave-vector projector 
$ P_T^{ij}=\delta^{ij}-\frac{k^ik^j}{k^2}$, 
the momentum can be split into the longitudinal 
and transverse parts.
The continuity equation 
in Eq.~\eqref{eq:exact} gives $ik_i\pi^i=-\partial_\tau\delta$ in Fourier space, thereby
completely 
fixing the 
longitudinal part, 
e.g. at the 2nd order we have\footnote{We now use the integrand shorthand notation \be 
 \int_{\q_1\q_2}^\k\dots\equiv\int\!\frac{d^3q_1}{(2\pi)^3}\frac{d^3q_2}{(2\pi)^3}(2\pi)^3
 \delta^{(3)}_D(\k-\q_1-\q_2).
\ee}
\be
 \left[\pi^i_{(2), L}\right]^c=\frac{ik^i}{k^2}\,\partial_\tau
 \delta^c_{(2)}
 \;=\;\frac{ik^i}{k^2}\,\partial_\tau\!\int_{\q_1,\q_2}^{\k} F_2^{\rm ctr}\delta_{(1)}(\q_1)\delta_{(1)}(\q_2) \,.
 \label{eq:longpart}
\ee 
The 2nd order counterterms 
are sourced by the 
stress-tensor expanded up
to 2nd order,
\begin{equation}
\label{eq:tauNLO}
\begin{split}
 \tau^i_\textit{vis}&=
-c_s^{2}\partial^i\triangle\Phi
 +\frac{c_v^2}{\H f}\partial^i\triangle u
 \;
-\,c_1\partial^i\delta_{(1)}^2-c_2\partial^i s^2
 \;-\,c_3\,s^{ij}\partial_j\delta_{(1)}\\
\tau_\textit{vis} &=\partial_i \tau^i_\textit{vis}=-d^2\Delta \delta_{(2)}
 \;
-\,e_1\Delta\delta_{(1)}^2-e_2\Delta s^2
 \;-\,e_3\,\partial_i(s^{ij}\partial_j\delta_{(1)})\,,
 \end{split}
\end{equation}
where we used $d^2\equiv c_s^2+c_v^2$ and the mapping
\begin{equation}
 e_1=c_1-\tfrac4{21}c_v^2,
 \qquad
 e_2=c_2+\tfrac27c_v^2,
 \qquad
 e_3=c_3 \,.
 \label{eq:credefVI}
\end{equation}
Using the Green's functions
for the density, one can 
obtain 
$F_2^{\text{ctr}}$,
which has been derived 
in~\cite{Baldauf:2014qfa},
\begin{align}\label{eq:F_2_ctr}
         F_2^{\text{ctr}}(\q,\k-\q,\tau)\,
    &= -\sum_{i=1}^{3} \epsilon_i\, E_i(\q,\k-\q)
       - \gamma \left[ \bar{F}_2^{\alpha\beta}(\q,\k-\q,\tau)
       + \bar{F}_2^{\delta}(\q,\k-\q,\tau) \right],
\end{align}
where the $\epsilon_i$ are time-dependent counterterms, while the other kernels are given explicitly in Equation \eqref{eq:E123}, and
$F_2^\delta=R_\delta k^2F_2^{\rm SPT}$. The kernel
$\bar F_2^{\alpha\beta}$
is defined using the 
fluid kernels $\alpha_{12}$, $\beta_{12}$:
\begin{align}
 \bar F_2^{\alpha\beta}
 &=\widetilde R_\alpha\big(q_1^2\alpha_{12}+q_2^2\alpha_{21}\big)
 +R_\alpha\big(q_1^2\alpha_{21}+q_2^2\alpha_{12}\big)
 +R_\beta(q_1^2+q_2^2)\beta_{12} \text{, with}\,\notag\\
 \alpha_{12}&=\frac{\q_1\cdot(\q_1+\q_2)}{q_1^2}, \beta_{12}=\frac{|\q_1+\q_2|^2\,(\q_1\cdot\q_2)}{2q_1^2q_2^2} 
 \label{eq:Fab} 
\end{align}
The time-dependent 
coefficients are given by 
\begin{equation}
\begin{split} 
\label{eq:defsVI}
&
 R_\alpha=\frac{g_2^\alpha}{g_1},
 \quad
 \widetilde R_\alpha=\frac{\widetilde g_2^\alpha}{g_1},
 \quad
 R_\beta=\frac{g_2^\beta}{g_1},
 \quad
 R_\delta=\frac{g_2^{d^2}}{g_1} \,,\\
 & g_2^{\alpha}\equiv-\frac{1}{2D_1^2}\!\int\! da'\,G_\delta\,
 \H\partial_{a'}\big\{a'\H f\,g_1D_1^2\big\},
 \qquad
 \widetilde g_2^{\alpha}\equiv-\frac{1}{2D_1^2}\!\int\! da'\,G_\delta\,
 \H\partial_{a'}\big\{a'\H f\,h_1D_1^2\big\}\,,
 \\
& g_1 \equiv\frac{1}{D_1}\int\! da'\,G_\delta\,d^2D_1\,, \qquad 
 g_2^{\beta}\equiv-\frac{1}{D_1^2}\!\int\! da'\,G_\delta\,\big[\H f\big]^2h_1D_1^2 \,,
\end{split}
\end{equation}
where $G_\delta(a,a')$
is the density Green's function of the fluid equations. 
The velocity divergence counterterm $h_1$ ($\theta^c_{(1)}=h_1k^2\delta_{(1)}$) is defined through
the continuity equation 
$a\partial_a(g_1D_1)=f h_1D_1$.
Note that $g_1$ is exactly minus
the dark matter sound speed counterterm
$\gamma=-g_1$. Moving forward it will
be useful to define 
\be 
\begin{aligned}
a_4&=2R_\alpha+R_\beta, \quad a_3&=-4R_\alpha,\quad 
 a_2&=-\tfrac27a_4, &a_1&=\widetilde R_\alpha-\tfrac97R_\alpha-\tfrac{17}{21}R_\beta \,.
 \end{aligned}
\ee 
Assuming an EdS scaling for 
$c_s^2+c_v^2=d^2\propto D_+^{m+1}$
the above coefficients
read
\begin{equation}
\begin{split}
& R_\alpha=\frac{2m+7}{2(m+2)(2m+9)},
 \qquad
 \widetilde R_\alpha=(m+2)R_\alpha,
 \qquad
 R_\beta=\frac{2}{2m+9},
\qquad 
 R_\delta=\frac{(m+1)(2m+7)}{(m+2)(2m+9)} \,,
\\
& a_1=\frac{42m^2+109m-31}{42(m+2)(2m+9)}, \quad 
 a_2=-\frac{2(4m+11)}{7(m+2)(2m+9)}, \quad 
 a_3=-\frac{2(2m+7)}{(m+2)(2m+9)}\,.
 \end{split}
\end{equation}
Introducing $E_4\equiv k^2F_2^{\rm SPT}(\q_1,\q_2)$,
it is easy to show 
that 
\be 
 \bar F_2^{\alpha\beta}
 +
 \bar F_2^{\delta}
 =a_1E_1+a_2E_2+a_3E_3+E_4 \,.
\ee 
Crucially, the coefficient in
front of $E_4$ is unity. 
This result is 
dictated by the equivalence 
principle and holds beyond
EdS or $d^2\propto D_+^{m+1}$
assumptions. This implies that 
\textit{without loss of generality} 
\be 
F_2^{\text{ctr}}(\q_1,\q_2)
= -\sum_{i=1}^{3} \tilde{\epsilon}_i\, 
E_i(\q_1,\q_2)
       - \gamma k^2 F_2^{\rm SPT}(\q_1,\q_2)\,,
\ee 
where $\gamma$ is the exact 
sound speed counterterm entering
the one-loop matter power spectrum, $P_{\rm 1-loop}^{\rm EFT}\supset -2k^2\gamma P_L(k)$.
The explicit
mapping
between the $\tilde{\epsilon}_A$
coefficients and 
the stress-tensor ones is $\tilde{\epsilon}_A=\epsilon_A+\gamma a_A$, 
\be 
 \epsilon_A\equiv-\frac{1}{D_1^2}\int\! da'\,G_\delta(a,a')\,X(a')D_1^2(a'),
 \qquad X\in\{d^2,e_1,e_2,e_3\} \,.
 \label{eq:g2X}
\ee 
Assuming 
$c_X(a)=\H_0^2 \Omega^0_m D_1^m(a)\,\bar c_X$,
the integrals over Green's functions evaluate to 
\begin{align}
 \gamma&=\frac{2\,\bar d^{\,2}}{(m+1)(2m+7)}\,D_1^{m+1},
 \qquad\qquad
 \epsilon_A=\frac{2\,\bar e_A}{(m+2)(2m+9)}\,D_1^{m+1},
 \nonumber\\[1ex]
 \widetilde\epsilon_1&=\frac{2D_1^{m+1}}{(m+2)(2m+9)}
 \left[\bar e_1+\frac{\big(42m^2+109m-31\big)\,\bar d^{\,2}}{42(m+1)(2m+7)}\right],
 \nonumber\\[0.5ex]
 \widetilde\epsilon_2&=\frac{2D_1^{m+1}}{(m+2)(2m+9)}
 \left[\bar e_2-\frac{2(4m+11)\,\bar d^{\,2}}{7(m+1)(2m+7)}\right],
 \label{eq:explicitVI}\notag\\[0.5ex]
 \widetilde\epsilon_3&=\frac{2D_1^{m+1}}{(m+2)(2m+9)}
 \left[\bar c_3-\frac{2\,\bar d^{\,2}}{m+1}\right]\,.
\end{align}
Let us derive now 
the transverse 
contribution.
It has two sources:
\be \left[\pi^i\right]^c_{T, (2)}=\big[v^i\big]^c_{T, (2)}+\big[\delta\,v^i\big]^c_{T, (2)} \,.
\ee
The first term
above 
is generated by the 
stress-tensor feedback
from
the rightmost term
in Eq.~\eqref{eq:tauNLO},
which is the only one that is not 
a total gradient.
The contact term $[\delta v^i]_T$
corresponds to renormalization of the mass weighted velocity field. 
The first term 
simply follows 
from the stress tensor
\be 
\big[v^i\big]^c_{T, (2)}(\tau)=\frac{1}{a(\tau)}\int^\tau\!d\tau'\;a(\tau')\,
 \big[\tau^i_\textit{vis}(\tau')\big]_T \,,\quad \big[\tau^i_\textit{vis}\big]_T= -
c_3\Big[s^{ij}\partial_j\delta\Big]_T\,,
\ee 
where we took into account that only
the $c_3$ contribution
can source a transverse
component, and the l.h.s. of the above equation features Green's function
of the vorticity component 
of the velocity field in the Euler equation. 
A direct computation gives
\begin{equation}
 \begin{split}
& \big[v^i\big]^c_{T, (2)}\;\equiv\;\frac{if\H}{2}\,c_\omega^2\int_{\q_1\q_2}^\k(\q_1\!\cdot\!\q_2)
 \Big(\frac1{q_1^2}-\frac1{q_2^2}\Big)n^i(\q_1, \q_2)\,\delta_{(1)}(\q_1)\delta_{(1)}(\q_2),
 \\
& c_\omega^2(a)\equiv g_\omega(a)
 =-\frac{1}{\H f\,a\,D_1^2}\int^a\!\frac{da'}{\H(a')}\,c_3(a')\,D_1^2(a') 
 \,,
 \label{eq:comegadef}
 \end{split}
\end{equation}
where the vector in the $(\q_1,\q_2)$
plane transverse to $\k$ is
\be 
n^i(\q_1, \q_2) = -\frac{q_2^i(\k
  \cdot \q_1)-q_1^i(\k\cdot
  \q_2)}{k^2}\,.
\ee 
Assuming
$c_3(a)=\H_0^2 \Omega^0_m D_1^{m+1}(a)\bar c_3$,
we get
$c_\omega^2=-\frac{2D_1^{m+1}}{2m+7}\,\bar c_3 $.
As for the second term, at second order we get a decomposition
\begin{equation}
 \big[\delta\,v^i\big]^c_{T, (2)}=\left[\delta_{(1)}[v^{i}]^c_{(1)}+\delta^c_{(1)}v^i_{(1)}\right]_T.
\end{equation}
The $\delta_{(1)}[v^{i}]^c_{(1)}$ term 
is longitudinal
and hence it drops from the transverse part. 
Using the first order solution $\delta^c_{(1)}=-\gamma k^2\delta_{(1)}$,
$v^i=i\H f\,\tfrac{q^i}{q^2}\,\theta$ 
and projecting the result onto the 
transverse direction
we get 
\be 
\big[v^i\big]^c_{T, (2)}
 =-\frac{if\H}{2}\,\gamma\int_{\q_1\q_2}^{\k}
 \Big(\frac{q_1^2}{q_2^2}-\frac{q_2^2}{q_1^2}\Big)
n^i(\q_1, \q_2)\,\delta_{(1)}(\q_1)
 \delta_{(1)}(\q_2).
 \label{eq:dvT}
\ee
Putting this together we arrive at Eq.~\eqref{eq:M_2_ctr}.
Note that the transverse part of this operator has one contribution stemming from the transverse velocity field, 
and another contribution 
from the transverse momentum field. The latter is non-zero even in SPT even though 
the velocity field is always longitudinal there. Indeed, applying the transverse projector $P^T_{ij}$
to the SPT kernel $M_2$ we obtain
\be 
[M_2^{i}]^T_{\rm SPT}(\q_1, \q_2) =
-\frac{if\HH}{2}
\frac{q_2^i (\q_{12}\cdot \q_1) - q_1^i(\q_{12}\cdot \q_2)}{q_{12}^2} \left[ \frac{1}{q_1^2} - \frac{1}{q_2^2}\right] ~\,.
\ee 

Let us now discuss
how much freedom do we have to model the coefficients in
$M_2^{\rm ctr}$
and $F_2^{\rm ctr}$
simultaneously. 
In the most general case when the time-dependence of $c_i$
is completely
unknown, 
we have 
$\gamma$, 
$\tilde{\epsilon}_{1,2,3}$
entering 
$F_2^{\rm ctr}$,
their time derivatives entering 
the $M_2^{\rm ctr}$ counterterm, 
plus $c_w^2$,
given by a time integral over $c_3$
different from the one 
entering the density counterterm coefficient.
This amounts to 9 fitting parameters in total. 
Assuming the EdS scaling $e_i=\H^2_0\Omega_m^0 D_1^{m}\bar e_i$, this number can be reduced to just four, if $m$ is fixed using arguments such as UV-scaling ($m=1$) or the Lifshitz symmetry ($m=5/3$): 
\be 
\begin{split}
 F_2^{c}&=-D_1^{m+1}\Big[\bar{\widetilde\epsilon}_1E_1+\bar{\widetilde\epsilon}_2E_2
 +\bar{\widetilde\epsilon}_3E_3+\bar\gamma\,k^2F_2\Big],
 \\[1.2ex]
 M_2^{i,\rm ctr}(\q_1, \q_2)&=i\,(m+3)f\H\,\frac{k^i}{k^2}\,F_2^{c}(\q_1, \q_2)
 \\[0.4ex]
 &\quad-\frac{i f\H\,D_1^{m+1}}{2}\,
 \frac{q_2^i(\bm q_{12}\!\cdot\!\q_1)-q_1^i(\bm q_{12}\!\cdot\!\q_2)}{q_{12}^2}
 \left[\bar c_\omega^2\,(\q_1\cdot\q_2)\Big(\frac1{q_1^2}-\frac1{q_2^2}\Big)
 +\bar\gamma\Big(\frac{q_1^2}{q_2^2}-\frac{q_2^2}{q_1^2}\Big)\right]\,,
\end{split}
\ee 
with 
\begin{equation}
\begin{aligned}\label{eq:ctr_time_dep_m}
 \bar{\widetilde\epsilon}_1&=\frac{1}{(m+2)(2m+9)}
 \left[2\,\bar e_1+\frac{42m^2+109m-31}{42}\,\bar\gamma\right],
 \\[0.6ex]
 \bar{\widetilde\epsilon}_2&=\frac{1}{(m+2)(2m+9)}
 \left[2\,\bar e_2-\frac{2(4m+11)}{7}\,\bar\gamma\right],
 \\[0.6ex]
 \bar{\widetilde\epsilon}_3&=\frac{2}{(m+2)(2m+9)}
 \Big[\bar e_3-(2m+7)\,\bar\gamma\Big],
 \qquad
 \bar c_\omega^2=-\frac{2\,\bar e_3}{2m+7}.
\end{aligned}
\end{equation}

\section{Bispectrum UV limits\label{app:bispec_uv_cancel}}
Here we present the explicit form of UV limits of the loop corrections $B_{321}^{II}$ and $B_{411}$ to demonstrate their cancellation by the counterterm corrections derived above. Instead of writing the entire expressions, we isolate the term inside the integral --- it can be seen between Eq. \eqref{eq:bispec_loops} and Eq. \eqref{eq:bispec_ctr_structure} that the prefactors to integrals and the counterterm graphs are identical. Starting with $B_{321}^{II}$, we have the leading UV limits of:
\begin{align}
    \lim_{q\gg k}\int_{\q}M_3(-\k_3, \q, -\q) P_{11}(q) &= \frac{61}{105}k_3 \mu_3 \int_{\q, q \gg k} \frac{P_{11}(q)}{q^2}\notag\\
    \lim_{q\gg k}\int_{\q}F_3(-\k_1, \q, -\q) P_{11}(q) &= -\frac{61}{315}k_1^2\int_{\q, q \gg k} \frac{P_{11}(q)}{q^2}.
\end{align}
These are renormalized by the $B_{c21}$ counterterm corrections, analogous to the case of the one-loop power spectrum,
\be \label{eq:Pkren}
\gamma=-\frac{61}{630}D^2\sigma^2_d(z=0)\equiv -\frac{61}{630}D^2\int_\q \frac{P_{11}(q,z=0)}{q^2}\,.
\ee 
This is expected, as these integrals appearing in $B_{321}^{II}$ are identical to the $P_{13}-$type corrections to the one-loop power spectrum. Continuing with $B_{411}$ in a similar fashion, we have: 
\begin{align}
    &\lim_{q\gg k}\int_{\q}M_4(\k_1, \k_2,\q, -\q) P_{11}(q) = \frac{F_s\,(k_1\mu_1 + k_2\mu_2) + F_a\,(k_1\mu_1 - k_2\mu_2)}{679140\,k_1^2 k_2^2 k_3^2}\int_{\q, q \gg k} \frac{P_{11}(q)}{q^2},\notag\\
    &\lim_{q\gg k}\int_{\q}F_4(\k_1, \k_3,\q, -\q) P_{11}(q) = -\frac{ G}{679140\, k_1^2 k_3^2}\int_{\q, q \gg k} \frac{P_{11}(q)}{q^2}, \text{ where}\notag\\
    &\quad F_s = 4008\,(k_1^2 - k_2^2)^2 (k_1^2 + k_2^2)
     - \big[\,101105\,(k_1^4 + k_2^4) - 152802\,k_1^2 k_2^2\,\big]\,k_3^2 \notag \\
        &\qquad + 80340\,(k_1^2 + k_2^2)\,k_3^4 + 49636\,k_3^6, \notag\\[4pt]
    &\quad F_a = 11\,(k_1^2 - k_2^2)\,k_3^2\,\big[\,4008\,(k_1^2 + k_2^2) - 6997\,k_3^2\,\big],\notag\\
    &\quad G = 12024\,(k_1^2 - k_3^2)^2 (k_1^2 + k_3^2)
   - \big[\,44518\,(k_1^4 + k_3^4) - 76684\,k_1^2 k_3^2\,\big]\,k_2^2 \notag \\
  &\qquad + 20085\,(k_1^2 + k_3^2)\,k_2^4 + 12409\,k_2^6.
\end{align}

We check that the counterterm expressions in Appendix \ref{app:M_n_ctr} are sufficient to renormalize these UV divergences appearing in the loop integrals. Additionally, in the case of an assumed power law time-dependence, a consistency relation can be established under the requirement that the UV divergences at leading order are exactly canceled, expressing all counterterms as functions of $\gamma$ and $m$.
For example, to cancel the UV divergence of $F_4$ by $F_2^{\text{ctr}}$, if we fix $m=1$, we obtain
\begin{align}
    \tilde{\epsilon}_1 = \frac{142}{427}\gamma, \; \tilde{\epsilon}_2=\frac{3015}{32879}\gamma, \; \tilde{\epsilon}_3=\frac{24048}{32879}\gamma~\,,
\end{align}
with the identification~\eqref{eq:Pkren}.
The above conditions exactly match \cite{Baldauf:2014qfa} after the transformation~\eqref{eq:ctr_time_dep_m}. To cancel the UV divergence of $M_4$ by $M_2^{\text{ctr}}$, we obtain an identical relation between the longitudinal momentum field counterterms with an extra factor of $f \HH$, as expected by the continuity equation. We also see that these pure longitudinal components are insufficient to renormalize the entire divergence, further motivating the necessity of the transverse counterterm expressions. Under the assumption of power law time dependence and for $m=1$, we obtain the relation 
\begin{align}
    c_\omega^2=-\frac{13994}{2989}\gamma.
\end{align}

\section{Bispectrum Gaussian Covariance \label{app:sigma_B}}
Here we present explicit expressions for the diagonal Gaussian covariance of the $B_{\delta \delta\pi_z}^{\ell=1}$ estimator. These expressions largely follow known results for the bispectrum multipole covariance, but with extra detail paid to the presence of the different $\delta$ and $\pi_z$ fields. Starting with the expression for the bispectrum estimator, Equation \eqref{eq:bispec_estimate_sum}, we have: 
\begin{align}
\big\langle \hat{B}_\ell\,\hat{B}_\ell \big\rangle(k_1, k_2, k_3)
 &= \left(\frac{2\ell+1}{2}\right)^{\!2}\frac{1}{V^2}\frac{1}{\mathcal{N}(k_1, k_2, k_3)^2}
 \notag\\&\times\sum_{\substack{\q_1+\q_2+\q_3=0\\ \q_1'+\q_2'+\q_3'=0}}\mathcal{L}_\ell(\z\cdot\hat\q_3)\,\mathcal{L}_\ell(\z\cdot\hat\q_3')\prod_{i=1}^3 \Theta_{k_i}(\q_i)\,\Theta_{k_i}(\q_i')\notag\\&
 \qquad\quad\times
  \big\langle \delta_m(\q_1)\delta_m(\q_2)\pi_z(\q_3)\,
             \delta_m(\q_1')\delta_m(\q_2')\pi_z(\q_3')\big\rangle.
\end{align}
 Taking care to Wick expand the correlation function for cases of special triangles, we obtain the following results: 
\begin{align}
\big\langle \hat{\tilde{B}}^2(k_1,k_2,k_3)\big\rangle_{\rm scalene}
 &= \left(\frac{2\ell+1}{2}\right)^{\!2}_{\ell=1}
 \frac{8\pi^4}{5}\;
 \frac{P_{11}(k_1)\,P_{11}(k_2)\,P_{11}(k_3)}
      {V\;k_1k_2k_3^3\;\Delta k^3}\\
\big\langle \hat{\tilde{B}}^2(k_1,k_1,k_3)\big\rangle_{\rm isosceles}
 &= \left(\frac{2\ell+1}{2}\right)^{\!2}_{\ell=1}
 \frac{16\pi^4}{5}\;
 \frac{[P_{11}(k_1)]^2\,P_{11}(k_3)}
      {V\;k_1^2k_3^3\;\Delta k^3}\\
\big\langle \hat{\tilde{B}}^2(k_1,k_2,k_1)\big\rangle_{\rm isosceles}
 &= \left(\frac{2\ell+1}{2}\right)^{\!2}_{\ell=1}
 \frac{8\pi^4}{5}
 \left[\frac{14}{9}+\frac{4}{9}\,\mathcal{L}_2(\Delta_{13})\right]
 \frac{[P_{11}(k_1)]^2\,P_{11}(k_2)}
      {V\;k_1^4k_2\;\Delta k^3}\\
\big\langle \hat{\tilde{B}}^2(k,k,k)\big\rangle_{\rm equilateral}
 &= \left(\frac{2\ell+1}{2}\right)^{\!2}_{\ell=1}
 \frac{32\pi^4}{5}\;
 \frac{[P_{11}(k)]^3}
      {V\;k^5\;\Delta k^3}
 \label{eq:summequil}
\end{align}
 with $\Delta_{13} = (k_1^2+k_3^2-k_2^2)/(2k_1k_3)$ at $k_3=k_1$, and the $k_2=k_3$ isosceles case given simply under a permutation of $k_1 \leftrightarrow k_2$. We note the novel angular structure of the isosceles cases, with the covariance depending explicitly on an angle of the triangle. We also note that the equilateral configuration acquires an effective symmetry factor of 4, as opposed to the usual case of an auto-bispectrum where this symmetry factor is 6, since we have a differentiated momentum field. In simplifying expressions, we have also substituted $\mathcal{N}(k_1, k_2, k_3)$ using a thin-bin approximation:
\begin{equation}
\mathcal{N}(k_1,k_2,k_3) = \frac{V^2}{(2\pi)^6}\,8\pi^2\,k_1k_2k_3\,\Delta k^3.
\end{equation}

\bibliographystyle{JHEP.bst}
\bibliography{kszeft.bib}

\end{document}